\documentclass[twocolumn,
showpacs,
showkeys,
preprintnumbers,
nofootinbib,
superscriptaddress,
amsmath,
amssymb,
floatfix,
secnumarabic,
aps,
pra,
a4paper,
notitlepage,
final,
]{revtex4-1}%

\usepackage[colorlinks=true,urlcolor=blue]{hyperref}
\usepackage{graphicx}
\usepackage{epsfig}
\usepackage{float}

\newcommand{\beq}{\begin{equation}}
\newcommand{\eeq}{\end{equation}}
\newcommand{\beqar}{\begin{eqnarray}}
\newcommand{\eeqar}{\end{eqnarray}}
\newcommand{\bal}{\begin{aligned}}
\newcommand{\eal}{\end{aligned}}

\def\ham{\hbox{$\cal H$}}
\def\dalam{\hbox
{\vrule\vbox{\hrule\hbox to 1ex{ \hfill}\kern 1 ex\hrule}\vrule}}

\def\1/2{\hbox{$ {1 \over 2}$ }}
\def\tr{\hbox{Tr}}

\def\h{\hbar}
\def\i/h{{i \over \h}}

\def\inf{\infty}
 
\def\v{\vec}

\def\a{\alpha}  
\def\b{\beta}  
 
\def\g{\gamma}  
\def\d{\delta} \def\D{\Delta} 
\def\l{\lambda}  
\def\e{\epsilon} \def\E{\hbox{$\cal E $}}

\def\s{\sigma}\def\S{\Sigma}
\def\r{\rho} \def\vr{\varrho}

\def\p{\psi} 
\def\bp{\bar \psi}

\def\m{\mu}
\def\n{\nu}
\def\t{\tau}

\def\z{\zeta}

\def\tt{\theta} \def\vt{\vartheta}

\def\<{\langle}
\def\>{\rangle}

\def\({\left(}
\def\[{\left[}
\def\){\right)}
\def\]{\right]}

\usepackage{subfigure}

\usepackage[notcite,notref,color]{showkeys}  

\usepackage{tikz}
\usetikzlibrary{decorations.pathreplacing}
\usetikzlibrary{decorations.pathmorphing}    
\usepackage{multirow}
\usepackage{dcolumn}		
\newcolumntype{.}{D{.}{.}{-1}}
\newcolumntype{i}[1]{D{.}{.}{#1}}

\newcommand{\myfrac}[2]{{\ifmmode{}^{#1}\!/_{\!#2}\else${}^{#1}\!/_{\!#2}$\fi}}

\begin{document}
\sloppy

\title{Lepton Number vs Coulomb Super-Criticality}

\author{A.Krasnov and K.Sveshnikov}
\email{aa.krasnov@physics.msu.ru}
\email{k.sveshnikov@physics.msu.ru}
\affiliation{Department of Physics and
Institute of Theoretical Problems of MicroWorld, Moscow State
University, 119991, Leninsky Gory, Moscow, Russia}


\date{\today}


\begin{abstract}
 The non-perturbative effects  of QED-vacuum reconstruction under influence of a supercritical EM-source ($Z>Z_{cr,1}$)  are considered in view of lepton number conservation. The question is here that if  the vacuum  shells formation and spontaneous positron emission, associated with discrete levels diving into the lower continuum, exist in reality, then it by no means should be a signal of new physics. The reason is that  the emitted positrons carry away the lepton number equal to $(-1)\times$their total number,  and so the   corresponding amount of positive lepton number should be shared by VP-density, concentrated in vacuum shells. In this case instead of integer lepton number of real particles there should appear the lepton number VP-density. Otherwise either the lepton number conservation  in such processes must be broken, or the spontaneous positron emission  prohibited.
 The conditions, under which the emission of vacuum positrons  can be unambiguously detected on the nuclear conversion pairs background, are also discussed.
\end{abstract}


\maketitle

\section{Introduction}

Nowadays the behavior of QED-vacuum under influence of a supercritical EM-source is subject to an active research~\cite{Rafelski2016,Kuleshov2015a,*Kuleshov2015b,*Godunov2017,Davydov2017,*Sveshnikov2017,
*Voronina2017,Popov2018,*Novak2018,*Maltsev2018,Roenko2018,Maltsev2019,*Maltsev2020}. Of the main interest is that  in such external fields there should take place  a deep vacuum state reconstruction, caused by discrete levels diving into the lower continuum and accompanied by such nontrivial effects as spontaneous positron emission combined with vacuum  shells formation (see e.g., Refs.~\cite{Greiner1985a,Plunien1986,Greiner2012,Ruffini2010,Rafelski2016} and citations therein). In 3+1 QED  such effects are expected for extended Coulomb sources of nucleus size with charges $Z>Z_{cr,1} \simeq 170$, which are large enough for direct  observation and probably could be created in low energy heavy ions collisions  at new heavy ion facilities like  FAIR (Darmstadt), NICA  (Dubna), HIAF (Lanzhou).

The aim of the present paper is to figure out the main consequences of such non-perturbative VP-effects with emphasis on lepton number conservation. This question appears naturally since the most important result of discrete levels diving into the lower continuum beyond the threshold of super-criticality is the   vacuum  shells formation accompanied by spontaneous  positron emission. At the same time,  the emitted positrons carry away the lepton number equal to $(-1)\times$their total number,  and so the   corresponding amount of positive lepton number should be shared by VP-density, concentrated in vacuum shells. In this case instead of integer lepton number of real particles ($e$-, $\mu$-, $\t$-leptons, corresponding neutrinos and their antiparticles) there appears the lepton number VP-density. Otherwise either the lepton number conservation  in such processes must be broken, or the positron emission  prohibited. Therefore the reliable observation of vacuum positrons could shed light on the nature of lepton number, which as well as the baryon one, is so far just a number, but not a conserved charge, associated with certain type of symmetry. So the reasonable conditions, under which the vacuum positron emission can be unambiguously detected on the nuclear conversion pairs background, should play an exceptional role  in slow  ions collisions, aimed at the search of such events\footnote{In recent papers~\cite{Popov2018,*Novak2018,*Maltsev2018,Maltsev2019,*Maltsev2020} the nuclear conversion pairs are denoted as created by dynamical pair-production mechanism, while the vacuum positrons as coming from the spontaneous one. It should be noted, however, that these two processes are quite different, since the conversion  produces real $e^-$ from the lowest $K\, , L\, , \dots $  atomic shells and/or real $e^+e^-$ pairs, whereas the spontaneous one results in positron emission combined with vacuum  shells formation.}.

In the present paper this problem is explored within the  Dirac-Coulomb problem (DC) with external static or adiabatically slowly varying spherically-symmetric Coulomb potential, created by uniformly charged sphere
\beq
\label{1.5a}
V(r)=-Z \a\,\( {1 \over R(Z)}\, \tt(R(Z)-r)+ {1 \over r}\, \tt(r-R(Z)) \)  \ ,
\eeq
or  charged ball with
\begin{multline}\label{1.6}
V(r)=- Z \a\,\( {3\, R^2(Z) - r^2 \over 2\,R^3(Z) }\, \tt(R(Z)-r) \ +  \right. \\ \left.  + \ {1 \over r}\, \tt(r-R(Z)) \)  \ .
\end{multline}
Here and henceforth the relation between the radius  of the Coulomb source and its charge is given by
\beq
\label{1.8}
R(Z) \simeq 1.2\, (2.5\, |Z|)^{1/3} \ \hbox{fm} \ ,
\eeq
which roughly imitates  the size of super-heavy nucleus with charge $Z$. In what follows $R(Z)$ will be quite frequently denoted simply as $R$.

The non-stationary approach, based on the time-dependent picture created by two  heavy ions slowly moving along the classical Rutherford trajectories~\cite{Reinhardt1981,*Mueller1988,Ackad2008,Popov2018,*Novak2018,*Maltsev2018,
Maltsev2019,*Maltsev2020}, looks more attractive, since it imitates the realistic scenario of attaining the super-critical region in heavy ions collision. At the same time,  the VP-effects at short internuclear distances achieved in the monopole approximation are in rather good agreement with exact two-center ones~\cite{Reinhardt1981,*Mueller1988,Tupitsyn2010,Maltsev2019,*Maltsev2020} and lie always in between those for sphere and ball upon adjusting properly the coefficient in relation (\ref{1.8}), because they are very sensitive to the latter. The main advantage of time-dependent approach is $ab \ initio$ description  of  pairs production  caused by Coulomb excitations of nuclei, while in adiabatic picture the latter should be considered  as an additional component. However, it will be argued below that the actual threshold for  vacuum positrons detection on the conversion pairs background turns out to be not less than $Z^\ast \simeq 210$, which lies beyond the existing nowadays  opportunities in heavy ion collisions.

As in other works on this topic ~\cite{Wichmann1956,Gyulassy1975, McLerran1975a,*McLerran1975b,*McLerran1975c,Greiner1985a,Plunien1986,Greiner2012,Ruffini2010, Rafelski2016},  radiative corrections from virtual photons are neglected. Henceforth, if it is not stipulated separately, relativistic units  $\hbar=m_e=c=1$ and the standard representation of  Dirac matrices are used. Concrete calculations, illustrating the general picture, are performed for $\a=1/137.036$ by means of Computer Algebra Systems (such as Maple 21) to facilitate  the analytic calculations  and GNU Octave code for boosting the numerical work.

\section{Vacuum shells formation}

The most efficient non-perturbative evaluation of the  VP-charge density $\vr_{VP}(\vec{r})$ is based on the  Wichmann and Kroll (WK) approach ~\cite{Wichmann1956,Gyulassy1975,Mohr1998}. The starting point of the latter is the vacuum value
\begin{multline} \label{3.1}
\vr_{VP}(\vec{r})=-\frac{|e|}{2}\(\sum\limits_{\e_{n}<\e_{F}} \p_{n}(\vec{r})^{\dagger}\p_{n}(\vec{r}) \ - \right. \\ \left. - \ \sum\limits_{\e_{n}\geqslant \e_{F}} \p_{n}(\vec{r})^{\dagger}\p_{n}(\vec{r}) \) \ .
\end{multline}
 In (\ref{3.1}) $\e_F=-1$ is the Fermi level, which in such problems with strong Coulomb fields is chosen at the lower threshold, while $\e_{n}$ and $\p_n(\vec{r})$ are the eigenvalues and properly normalized set of  eigenfunctions of  corresponding DC. The expression (\ref{3.1}) for the  VP-charge density is a direct consequence of the well-known Schwinger prescription for the  fermionic current in terms of the fermion fields commutators
\beq
\label{3.1a}
 j_{\mu}(\v r, t)=-{|e| \over 2}\, \[ \bp(\v r, t)\,, \g_{\mu} \p(\v r, t)\] \ .
 \eeq

\begin{figure}
\center
\includegraphics[scale=0.20]{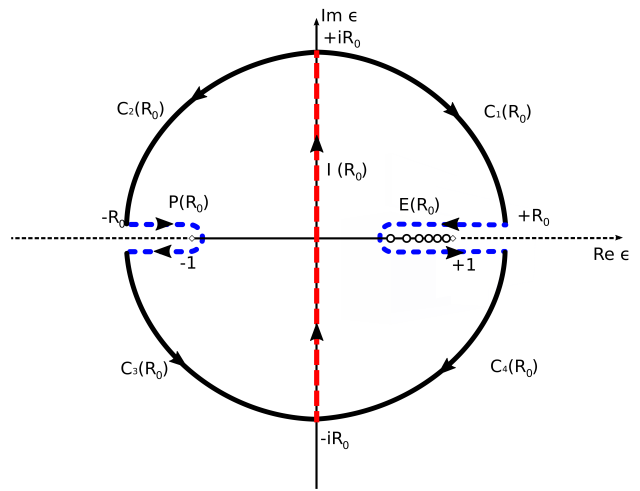} \\
\caption{\small (Color online) WK-contours in the complex energy plane, used for representation of the VP-charge density (\ref{3.1}) via contour integrals. The direction of contour integration is chosen in correspondence with (\ref{3.3}).}
\label{WK}
\end{figure}

The essence of the WK techniques is the representation of the density (\ref{3.1}) in terms of contour integrals on the first sheet of the Riemann energy plane containing the trace of the Green function of  corresponding DC. In our case the  Green function is defined via equation
\begin{multline}
\label{3.2}
\[-i \v{\a}\,\v{\nabla}_r +\b +V(r) -\e \]G(\vec{r},\vec{r}\,' ;\e) \\ =\d(\vec{r}-\vec{r}\,' ) \ .
\end{multline}
The formal solution of (\ref{3.2}) reads
\beq
\label{3.3}
G(\vec{r},\vec{r}\ ';\e)=\sum\limits_{n}\frac{\p_{n}(\vec{r})\p_{n}(\vec{r}\ ')^{\dagger}}{\e_{n}-\e} \ .
\eeq
Following Ref.~\cite{Wichmann1956},  the density (\ref{3.1}) is expressed via the integrals along the contours  $P(R_0)$ and $E(R_0)$ on the first sheet of energy surface  (Fig.\ref{WK}).
\begin{multline}
\label{3.4}
\vr_{VP}(\vec{r}) = \\ = -\frac{|e|}{2} \lim_{R_0\rightarrow \infty}\( \frac{1}{2\pi i}\int\limits_{P(R_0)} \! d\e\, \mathrm{Tr}G(\vec{r},\vec{r'};\e)|_{\vec{r'}\rightarrow \v r} \ + \right. \\ \left.+ \ \frac{1}{2\pi i}\int\limits_{E(R_0)} \! d\e\, \mathrm{Tr}G(\vec{r},\vec{r'};\e)|_{\vec{r'}\rightarrow \v r} \) \ .
\end{multline}
Note that the Green function in this relation must be properly regularized to insure that the limit $\vec {r'} \to \v r$ exists and that the integrals over $d\e$ converge. This regularization is discussed below. On this stage, though, all expressions are to be understood to involve only regulated Green functions. One of the main consequences of the last convention is the uniform asymptotics of the integrands in (\ref{3.4}) on the large circle $|\e| \to \inf$ at least as $O(1/\e)$, which allows for deforming the contours  $P(R_0)$ and $E(R_0)$ to the imaginary axis segment $I(R_0)$ and taking the limit $R_0 \to \inf$, what gives
\begin{multline}
\label{3.4a}
\vr_{VP}(\vec{r}) = \\ =|e| \[ \sum \limits_{-1 \leqslant \e_n < 0} |\p_n (\v r)|^2 +{1 \over 2\pi}\,  \int\limits_{-\inf}^{\inf} \! dy\, \mathrm{Tr}G(\vec{r},\vec{r'};iy)|_{\vec{r'}\rightarrow \v r}\]  \ ,
\end{multline}
where $\{\p_n(\v r)\}$ are the normalized eigenfunctions of negative discrete levels with $-1 \leqslant \e_n <0$ and here and henceforth $\p_{n}(\vec{r})^{\dagger}\p_{n}(\vec{r})=|\p_n (\v r)|^2$.

Proceeding further, the Green function (\ref{3.3}) is represented as a partial series over $k=\pm(j+1/2)$ ~\cite{Wichmann1956,Gyulassy1975}
\beq
\label{3.12}
\mathrm{Tr}G(\vec{r},\vec{r'};\e)|_{\vec{r'}\rightarrow \v r} = \sum\limits_k { |k| \over 2 \pi}\,\mathrm{Tr}G_k(r,r';\e)|_{r'\rightarrow r} \ ,
\eeq
where the radial Green function  $G_k(r,r';\e)$ is defined via
\beq
\label{3.13}
\ham_k(r)\,G_k(r,r';\e)={\d (r-r') \over r r'} ,
\eeq
while the  radial Dirac Hamiltonian takes the form
\beq
\label{3.13a}
\ham_k(r)=  \begin{pmatrix} V(r)+1-\e & -{1 \over r}\,{d \over dr}\,r - {k \over r} \\ {1 \over r}\,{d \over dr}\,r - {k \over r} & V(r)-1-\e  \end{pmatrix}  \ .
\eeq
 For the partial terms of $\vr_{VP}(r)$ one obtains
\begin{multline}
\label{3.14}
\vr_{VP,k}(r)  = {|e| |k| \over 2 \pi} \[ \sum \limits_{-1 \leqslant \e_{n,k} < 0} |\p_{n,k} (r)|^2 + \right. \\ \left. + {1 \over 2\pi}\,  \int\limits_{-\inf}^{\inf} \! dy\, \mathrm{Tr}G_k(r,r';iy)|_{r'\rightarrow r}\]  \ ,
\end{multline}
where $\p_{n,k} (r)$ are the normalized radial wave functions with eigenvalues $k$  and $\e_{n,k}$  of the corresponding DC. From (\ref{3.14}) by means of symmetry relations~\cite{Gyulassy1975}  for the sum $\vr_{VP,|k|}(r)$ of two partial VP-densities with opposite signs of $k$  one finds
\begin{multline}
\label{3.21}
\vr_{VP,|k|}(r)  =  {|e| |k| \over 2 \pi}\, \Big\{\sum \limits_{k=\pm |k|} \sum \limits_{-1 \leqslant \e_{n,k} < 0}  |\p_{n,k} (r)|^2  + \\ + {2 \over \pi}\,  \int\limits_{0}^{\inf} \! dy\, \mathrm{Re}\[\mathrm{Tr}G_{k}(r,r';iy)\]|_{r'\rightarrow r} \Big\} \ ,
\end{multline}
which is  by construction real and odd in $Z$ (in accordance with the Furry theorem).

The general result, obtained in Ref.~\cite{Gyulassy1975} within the expansion of $\vr_{VP}(\v r)$ in powers of  $Z\a$ (but with fixed radius $R=R(Z)$\,!), which is valid for 1+1 and 2+1 D always, and for a spherically-symmetric external potential in the three-dimensional case, is that all the divergencies of $\vr_{VP}(\v r)$ originate  from the fermionic loop with two external photon lines, whereas  the next-to-leading orders of expansion in $Z\a$ are already free from divergencies (see also Ref.~\cite{Mohr1998} and refs. therein). In the  non-perturbative approach this statement has been verified for 1+1 D in Refs.~\cite{Davydov2017,*Sveshnikov2017,
*Voronina2017} and for 2+1 D in Refs.~\cite{Davydov2018a,*Davydov2018b,Sveshnikov2019a,*Sveshnikov2019b} by means of the phase integral techniques. The latter can be also extended for the present 3+1 D case\footnote{We drop here all the intermediate steps, required for explicit construction of radial Green functions $G_{k}(r,r';\e)$ for the potentials (\ref{1.5a},\ref{1.6}) and justification of the limit $r' \to r$. For these steps one needs to deal with explicit solutions of DC and corresponding Wronskians, which takes a lot of space and so will be considered in a separate paper~\cite{Sveshnikov2022}.}. From these results there follows first that for any partial Green function $G_k$ the deformation of the WK-contours  $P(R_0)$ and $E(R_0)$ to the imaginary axis can be performed without any ambiguities. Second, the decreasing asymptotics of  $ \mathrm{Re}\,\[\tr G_{k} (r,r; i y)\]$ for $|y| \to \inf$ turns out to be much faster, namely $O(1/|y|^3)$, than the previously announced one for the regulated Green function $ G\(\v r,\v r'; \e\)$, which enters the initial WK-contour representation (\ref{3.4}). The main reason for such difference is that the properties of $G_k$  are much better than for the whole partial series. This circumstance has been discussed earlier in Ref.\cite{Gyulassy1975}.

So the renormalization procedure for the VP-density (\ref{3.1}) is actually the same for all the three spatial dimensions and reduces to the diagram with two external lines.
Thus, in order to find the renormalized   $\vr^{ren}_{VP}(r)$, the linear in external field terms in the expression  (\ref{3.4a}) should be extracted and replaced by the renormalized  first-order perturbative density $\vr^{(1)}_{VP}(r)$, evaluated with the same $R(Z)$. For these purposes let us introduce first the component  $\vr^{(3+)}_{VP,|k|}(r)$ of partial VP-density, which is defined in the next  way
\begin{multline}\label{3.25}
\vr_{VP,|k|}^{(3+)}(r)= \frac{|e||k|}{2 \pi}\,\Big\{ \sum\limits_{k=\pm|k|}\sum\limits_{-1\leqslant \e_{n,k}<0}|\p_{n,k}(r)|^2 \ +  \\  + \ \frac{2}{\pi} \int\limits_{0}^{\infty}d y\,\mathrm{Re}\[\tr G_{k}(r,r;iy)- \tr G^{(1)}_{k}(r;i y)\] \Big\} \ ,
\end{multline}
where $G^{(1)}_{k}(r;i y)$ is the  linear in $Z\a$ component of the partial Green function $G_{k}(r,r;i y)$ and so coincides with the first term of the Born series
\beq \label{3.26}
G^{(1)}_{k}=G^{(0)}_{k} (-V) G^{(0)}_{k} \ ,
\eeq
where $G_{k}^{(0)}$ is the free radial Green function with the same  $k$ and $\e$.
By construction $\vr^{(3+)}_{VP,|k|}(r)$ contains only  odd powers of $Z\a$ starting from $n=3$ and so is free of divergencies, but at the same time it is responsible for all the nonlinear effects, which are caused by   discrete levels  diving into the lower continuum.

As  a result, the renormalized  VP-charge density $\vr^{ren}_{VP}(r)$ is defined by the following expression\footnote{The convergence of the partial series in (\ref{3.27}) is shown explicitly for the point source in the original work by Wichmann and Kroll~\cite{Wichmann1956}, while accounting for the finite size of the source is discussed in detail in Ref.~\cite{Gyulassy1975}. For the present case it follows from convergence of the partial series for  $\E^{ren}_{VP}$, which is explicitly shown in~\cite{Sveshnikov2022}.}
\beq\label{3.27}
\vr^{ren}_{VP}(r)=\vr_{VP}^{(1)}(r)+\sum\limits_{k=1}^{\inf}\vr_{VP,|k|}^{(3+)}(r) \ ,
\eeq
where $\vr_{VP}^{(1)}(r)$ is the first-order perturbative VP-density, which is obtained from the one-loop (Uehling) potential $A^{(1)}_{VP,0}(\vec{r})$ in the next way ~\cite{Itzykson1980, Greiner2012}
\beq
\label{2.2}
\vr^{(1)}_{VP}(\vec{r})=-\frac{1}{4 \pi} \D A^{(1)}_{VP,0}(\vec{r}) \ ,
\eeq
where
\beq\begin{gathered}
\label{2.3}
A^{(1)}_{VP,0}(\vec{r})=\frac{1}{(2 \pi)^3} \int d \v q\,\, \mathrm{e}^{i \vec{q} \vec{r}}\, \Pi_{R}(-{\v q}^2)\,\widetilde{A}_{0}(\vec{q}) \ , \\
\widetilde{A}_{0}(\vec{q})=\int d \v r'\,\mathrm{e}^{-i \vec{q} \vec{r\,}' }\,A^{ext}_{0}(\vec{r}\,' ) \ .
\end{gathered}\eeq
The polarization function  $\Pi_R(q^2)$, which enters eq. (\ref{2.3}), is defined via general relation $\Pi_R^{\m\n}(q)=\(q^\m q^\n - g^{\m\n}q^2\)\Pi_R(q^2)$ and so is dimensionless. In the adiabatic case under consideration  $q^0=0$ and $\Pi_R(-{\v q}^2)$ takes the  form
\begin{multline}
\label{2.4}
\Pi_R(-{\v q}^2) = \\ = {2 \a \over \pi}\, \int \limits_0^1 \! d\b\,\b(1-\b)\,\ln\[1+\b(1-\b)\,{{\v q}^2 \over m^2-i\e}\] = \\ = {\a \over \pi}\, S\(|\v q|/m \) \ ,
\end{multline}
where
\begin{multline}\label{2.4a}
S(x)= -5/9 + 4/3 x^2 + (x^2- 2)\, \sqrt{x^2+4} \ \times \\ \times \ \ln \[ \(\sqrt{x^2+4}+x\) \Big/ \(\sqrt{x^2+4}-x\) \]/3x^3  \ .
\end{multline}

It would be worth to note that such expression for $\vr^{ren}_{VP}(r)$ provides that the  total VP-charge
\beq
\label{3.27a}
Q^{ren}_{VP}=\int\limits d {\v r}\,\vr^{ren}_{VP}(\v r)
\eeq
for $Z<Z_{cr,1}$ is zero, since  $Q_{VP}^{(1)}=0$ by construction, while the subsequent direct check confirms that the contribution of $\vr^{(3+)}_{VP}$ to  $Q^{ren}_{VP}$ for $Z<Z_{cr,1}$ vanishes too. In 1+1 D  such a check is implemented  purely analytically~\cite{Davydov2017}, while in 2+1 D due to complexity of  expressions, entering $\r^{(3+)}_{VP,|m_j|}(r)$, it requires a special combination of analytical and numerical methods (see ~\cite{Davydov2018a}, App.B). For 3+1 D this combination is extended with minimal changes, since the structure of partial Green functions in two- and three-dimensional cases is actually the same up to additional factor $1/r$ and replacement $m_j \to k$. Moreover, it suffices to verify that  $Q_{VP}^{ren}=0$ not for the whole subcritical region, but only in absence of negative discrete levels. In presence of the latters,  vanishing of the  total VP-charge for $Z<Z_{cr,1}$ can be figured out by means of the  model-independent arguments, which are based on the initial  expression for the vacuum density (\ref{3.1}). It follows from (\ref{3.1}) that the change of $Q_{VP}^{ren}$ is possible for  $Z>Z_{cr,1}$ only when the  discrete levels attain the lower continuum.   One of the possible correct ways to prove this statement is outlined in Ref.~\cite{Davydov2017}.  It should be specially mentioned that this effect is essentially non-perturbative and  completely included in $\vr_{VP}^{(3+)}$, while  $\vr_{VP}^{(1)}$ lies aside. Thus, the behavior of the renormalized via (\ref{3.27}) VP-density in the non-perturbative region turns out to be just the one that should be expected from general assumptions about the structure of the electron-positron vacuum for $Z> Z_{cr}$.

A more detailed picture of the changes in $\vr^{ren}_{VP}(r)$ for  $Z>Z_{cr,1}$ turns out to be quite similar to that considered in Refs.~\cite{Greiner1985a,Plunien1986,Greiner2012, Rafelski2016} by means of U.Fano approach to auto-ionization processes in atomic physics \cite{Greiner2012,Fano1961}. The main result is that, when the discrete level  $\p_{\n}(\v r)$ reaches the lower continuum, the change in the   VP-charge density equals to
\beq\label{3.28}
\D\vr_{VP}^{ren}(\v r)=-|e|\times |\p_{\n} (\v r)|^2 \ .
\eeq
Here it  should be noted first that the original approach \cite{Fano1961} deals directly with the change of density of states $n(\v r)$ and so the change in the induced charge density (\ref{3.28}) is  just a consequence of the basic relation $\vr(\v r)= -|e| n(\v r)$.  Furthermore, such a jump in VP-charge density occurs for each diving level with its unique set of quantum numbers $\{\n\}$. So in the present case each discrete $2|k|$-degenerated level upon diving into the lower continuum yields the jump of total VP-charge  equal to  $(-2|e||k|)$. The other quantities including the lepton number should reveal a similar behavior. Second, the Fano approach contains also a number of approximations. Actually, the expression (\ref{3.28}) is exact only in the  vicinity of the corresponding $Z_{cr}$, that  is shown by a number of concrete examples in Refs.~\cite{Davydov2017,*Sveshnikov2017,
*Voronina2017,Davydov2018a,Sveshnikov2019a,Voronina2019a}. So the most correct way to find $\vr_{VP}^{ren}(\v r)$ for the entire range of  $Z$ under study is to use the expression (\ref{3.27}) with subsequent direct check of the expected integer value of  $Q_{VP}^{ren}$.

\section{VP-energy in the non-perturbative approach}

The most consistent way to explore the spontaneous positron emission is to deal first with VP-energy $ \E_{VP} $, since there are indeed the changes in $ \E_{VP} $ caused by discrete levels diving, which  are responsible for creation of vacuum positrons. Actually, $ \E_{VP} $ is nothing else but the Casimir vacuum energy for the electron-positron system~\cite{Plunien1986}. The starting expression for $ \E_{VP} $ is
 \beq
\label{3.29a}
\E_{VP}=\frac{1}{2}\(\sum\limits_{\e_{n}<\e_{F}} \e_n  -   \sum\limits_{\e_{n}\geqslant \e_{F}} \e_n \) \ .
\eeq
The expression  (\ref{3.29a}) is obtained from the Dirac hamiltonian, written in the form which is  consistent with Schwinger prescription for the current (\ref{3.1a})  (for details see, e.g., Ref.~\cite{Plunien1986}) and is defined up to a constant, depending on the choice of the energy reference point. In (\ref{3.29a}) VP-energy is negative and divergent even in absence of external fields $ A_{ext} = 0 $. But since the VP-charge density is defined so that  it vanishes for $A_{ext} = 0 $, the natural choice of the reference point for $ \E_{VP} $ should be the same. Furthermore, in presence of the external Coulomb field of the type (\ref{1.5a},\ref{1.6}) there appears in the sum (\ref{3.29a}) also an (infinite) set of discrete levels. To pick out  exclusively the interaction effects it is therefore necessary to subtract from each discrete level the mass of the free electron at rest.

Thus, in the physically motivated form and in agreement with $\vr_{VP}$, the initial expression for the VP-energy should be written as
\begin{multline}
\label{3.29}
\E_{VP}=\1/2 \(\sum\limits_{\e_n<\e_F} \e_n-\sum\limits_{\e_n \geqslant \e_F} \e_n + \sum\limits_{-1\leqslant \e_n<1} \! 1
\)_A \ - \\ - \ \1/2 \(\sum\limits_{\e_n \leqslant -1} \e_n-\sum\limits_{\e_n
	\geqslant 1} \e_n \)_0 \ ,
\end{multline}
where the label A denotes the  non-vanishing external field $A_{ext}$, while the label 0 corresponds to   $A_{ext}=0$.  Defined in such a way,  VP-energy vanishes by turning off the external field, while by turning on it contains only the interaction effects, and so  the expansion of $ \E_{VP} $ in (even) powers of the external field starts from $ O\((Z\a)^2\) $.

Now let us extract  from (\ref{3.29}) separately the contributions from the discrete and continuous spectra for each value of angular quantum number $k$, and afterwards use for the difference of integrals over the continuous spectrum $ (\int d {\v q}\, \sqrt{q^2+1})_A-(\int d {\v q}\,\sqrt{q^2+1})_0 $ the well-known technique, which represents this difference in the form of an integral of the elastic scattering phase $ \d_k(q)$~\cite{Sveshnikov2017,Rajaraman1982,Sveshnikov1991,Jaffe2004}.  Omitting a number of almost obvious intermediate steps, which have been considered in detail in Ref.~\cite{Sveshnikov2017}, let us  write the final answer for $ \E_{VP}(Z) $ as a partial series
\beq
\label{3.30}
\E_{VP}(Z)=\sum\limits_{k=1}^{\inf} \E_{VP,k}(Z) \ ,
\eeq
where
\begin{multline}
\label{3.31}
\E_{VP,k}(Z)  = k\, \(\frac{1}{\pi} \int\limits_0^{\inf} \!   \  \frac{q \, dq }{\sqrt{q^2+1}} \ \d_{tot}(k,q) \ + \right. \\ \left. + \ \sum\limits_{\pm}\sum\limits_{-1 \leqslant \e_{n,\pm k}<1} \(1-\e_{n,\pm k}\)\) \ .
\end{multline}
In (\ref{3.31}) $ \d_{tot}(k,q) $ is the total phase shift for the given values of the wavenumber $q$ and angular number $\pm k$, including the  contributions from the scattering states from both  continua and  both parities for the radial DC problem with the  hamiltonian (\ref{3.13}). In the discrete spectrum contribution to $\E_{VP,k}(Z)$ the additional sum $\sum_{\pm}$ takes  also account of both parities.

Such approach to evaluation of $ \E_{VP} $ turns out to be quite effective, since for the external potentials of the type (\ref{1.5a},\ref{1.6})  each partial VP-energy turns out to be finite without any special regularization. First, $ \d_{tot}(k,q) $ behaves both in IR and UV-limits of the $q$-variable much better, than each of the scattering phase shifts separately.  Namely,  $ \d_{tot}(k,q) $ is finite for $ q \to 0 $ and behaves like  $ O(1/q^3) $ for $ q \to \inf $, hence, the phase integral in (\ref{3.31}) is always convergent. Moreover, $ \d_{tot}(k,q) $ is by construction an even function of the external field.  Second, in the bound states contribution to $ \E_{VP,k} $  the condensation point $ \e_{n,\pm k} \to 1 $ turns out to be regular, because  $ 1-\e_{n,\pm k} \sim O(1/n^2) $ for $ n \to \inf $. The latter circumstance permits to avoid intermediate cutoff of the Coulomb asymptotics of the external potential for $ r \to \inf $, what significantly simplifies all the subsequent calculations.

The principal problem of convergence of the partial series (\ref{3.30}) can be solved along the lines of Ref.~\cite{ Davydov2018b,Sveshnikov2019b}, demonstrating the convergence of the similar  expansion for $ \E_{VP} $ in 2+1 D. Due to a lot of details this topic is discussed separately~\cite{Sveshnikov2022}. The main result is that as expected from general grounds discussed above in terms of $\vr_{VP}(\v r)$ and in accordance  with similar results in 1+1 and 2+1 D cases ~\cite{Davydov2017,*Sveshnikov2017,
*Voronina2017,Davydov2018a,Davydov2018b,Sveshnikov2019a,Sveshnikov2019b},  the partial series (\ref{3.30})  for $\E_{VP}$ diverges quadratically in the leading $O((Z\a)^2)$-order. So it requires  regularization and subsequent renormalization, although each partial $\E_{VP,k}$  is finite without any additional manipulations. Moreover, the  divergence of the partial series (\ref{3.30}) is formally the same as in 3+1 QED  for the  fermionic loop with two external lines.

Thus, in the complete analogy with the renormalization of VP-density  (\ref{3.27}) we should pass to the renormalized VP-energy by means of  relations
\beq
\label{3.58}
\E^{ren}_{VP}(Z) =\sum\limits_{k=1} \E^{ren}_{VP,k}(Z) \ ,
\eeq
where
\beq
\label{3.59}
\E^{ren}_{VP,k}(Z)=\E_{VP,k}(Z)+ \z_k Z^2 \ ,
\eeq
while the renormalization coefficients $\z_k$ are defined in the following way
\beq
\label{3.60}
\z_k = \lim\limits_{Z_0 \to 0}  \[{\E_{VP}^{(1)}(Z_0)\,\d_{k,1}-\E_{VP,k}(Z_0) \over Z_0^2}\]_{R=R(Z)} \ . \eeq
The  essence of relations  (\ref{3.58}-\ref{3.60}) is to remove (for fixed  $R=R(Z)$\,!) the divergent $O((Z\a)^2)$-components from the non-renormalized partial terms  $ \E_{VP,k}(Z) $ in the series (\ref{3.30}) and  replace them further by renormalized via fermionic loop  perturbative contribution to VP-energy $\E^{(1)}_{VP}\,\d_{k,1}$. Such procedure provides simultaneously the convergence of the regulated this way partial series (\ref{3.58}) and the correct limit of $\E^{ren}_{VP}(Z)$  for $Z\a \to 0$ with fixed $R=R(Z)$.

The perturbative term $\E^{(1)}_{VP}\,\d_{k,1}$ is obtained from the general first-order relation
\beq
\label{2.1}
\E^{(1)}_{VP}=\frac{1}{2} \int d \v r\, \vr^{(1)}_{VP}(\vec{r})\,A_{0}^{ext}(\vec{r}) \ ,
\eeq
where $\vr^{(1)}_{VP}(\vec{r})$ is defined in (\ref{2.2}-\ref{2.4a}).
Proceeding further, from (\ref{2.1}) and (\ref{2.2},\ref{2.3}) one finds
\begin{multline}
\label{2.11}
\E^{(1)}_{VP}=\frac{1}{64 \pi^4}\, \int \! d \v q \ {\v q}^2\, \Pi_R(-{\v q}^2) \ \times \\ \times \  \Big| \int \! d \v r\, \mathrm{e}^{i \vec{q} \vec{r}}\,A_{0}^{ext}(\vec{r}) \Big|^2  \ .
\end{multline}
Note that since the function  $S(x)$ in (\ref{2.4}) is strictly positive, the perturbative VP-energy is positive too.

In the spherically-symmetric case  with $A_{0}^{ext}(\vec{r})=A_0(r)$  the perturbative VP-term belongs to the $s$-channel, which gives the factor $\d_{k,1}$ and
\begin{multline}
\label{2.12}
\E^{(1)}_{VP}=\frac{1}{\pi}\, \int\limits_0^{\inf} \! d q \  q^4\, \Pi_R(-q^2) \ \times \\ \times \  \( \int\limits_0^{\inf} \! r^2\, d r\, j_0(q r)\, A_{0}(r) \)^2  \ ,
\end{multline}
 whence for the sphere there follows
\beq
\label{2.15}
\E^{(1)}_{VP,\,sphere}={ (Z\a)^2  \over 2 \pi R}\, \int\limits_0^{\inf} \! {d q \over q} \ S(q/m) \ J_{1/2}^2(q R) \ ,
\eeq
while for the ball one obtains
\beq
\label{2.17}
\E^{(1)}_{VP,\,ball}={ 9\,  (Z\a)^2  \over 2 \pi R^3}\, \int\limits_0^{\inf} \! {d q \over q^3} \ S(q/m) \ J_{3/2}^2(q R) \ .
\eeq
By means of the condition
\beq
\label{2.18}
m R(Z) \ll 1 \ ,
\eeq
which is satisfied by the Coulomb source with relation (\ref{1.8}) between its charge and radius up to $Z \sim 1000$, the integrals (\ref{2.15},\ref{2.17}) can be  calculated analytically (see Ref.~\cite{Plunien1986} for details). In particular,
\beq\label{2.20}
\E^{(1)}_{VP,\,sphere}={ (Z\a)^2  \over 3 \pi R}\,  \[ \ln\( {1 \over 2 m R}\) - \g_E + {1 \over 6}\]
\eeq
for the sphere and
\beq
\label{2.21}
\E^{(1)}_{VP,\,ball}={2\, (Z\a)^2 \over 5 \pi R}\,  \[ \ln\( {1 \over 2 m R}\) - \g_E +  {1 \over 5} \]
\eeq
for the ball.

\section{The results of calculations}

For greater clarity  of results  we restrict this presentation to the case of charged sphere (\ref{1.5a}) on the interval $0 < Z < 600$ with the numerical coefficient in eq. (\ref{1.8}) chosen as\footnote{With such a choice for a charged ball with $Z=170$ the lowest $1s_{1/2}$-level lies precisely at $\e_{1s}=-0.99999$. Furthermore, it is quite close to $1.23$, which is the most commonly used coefficient in heavy nuclei physics.}
\beq
\label{6.1}
R(Z)=1.228935\, (2.5\, Z)^{1/3} \ \hbox{fm} \ .
\eeq
On this interval of $Z$  the main contribution comes from the partial channels with $k=1\,,2\,,3\,,4$, in which a non-zero number of discrete levels has already reached the lower continuum (for details see Ref.~\cite{Sveshnikov2022}).

\begin{figure*}[t!]
\subfigure[]{
		\includegraphics[width=1.31\columnwidth]{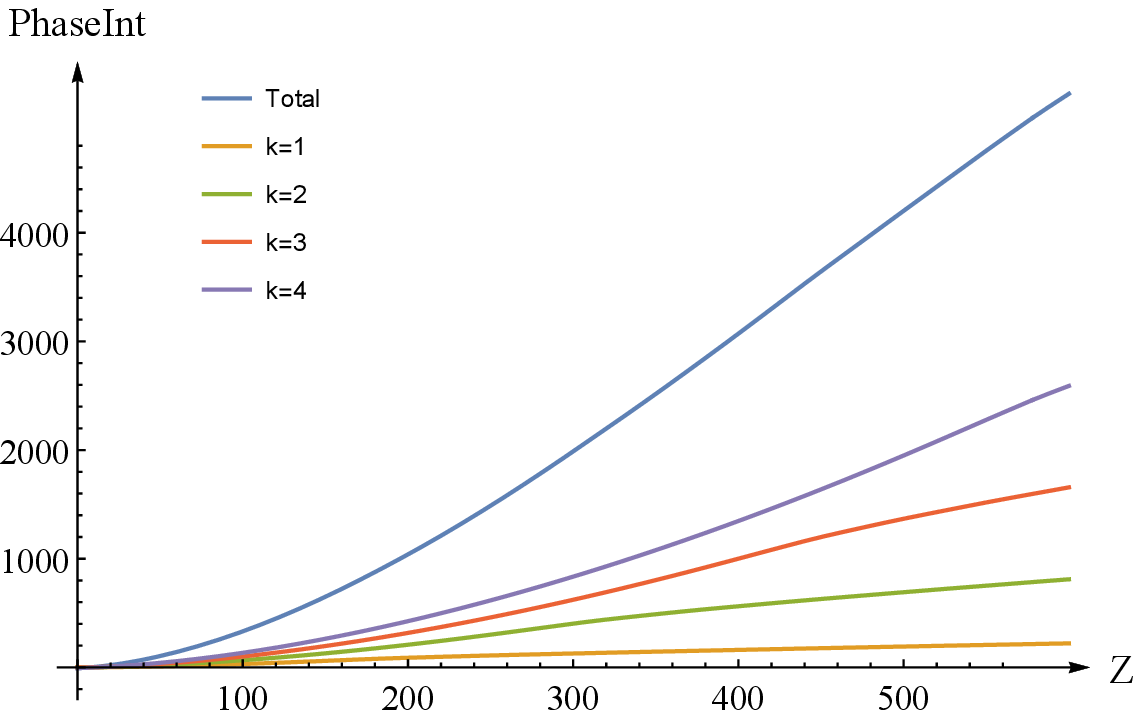}
}
\vfill
\subfigure[]{
		\includegraphics[width=1.31\columnwidth]{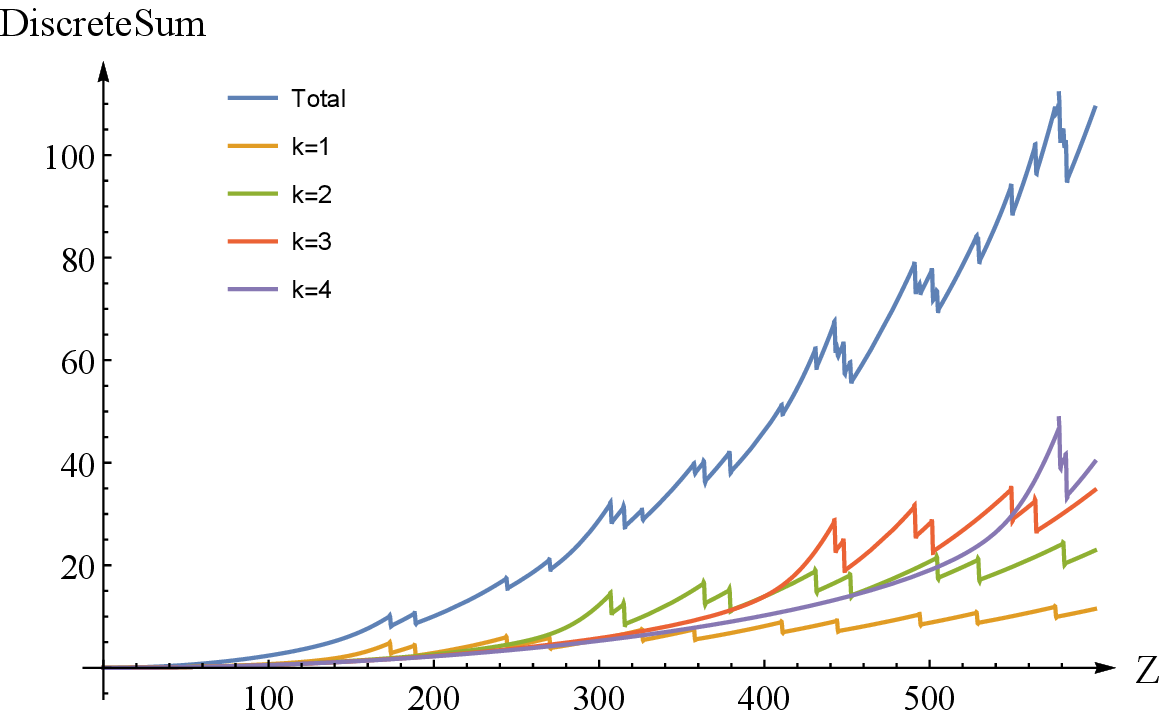}
}
\vfill
\subfigure[]{
		\includegraphics[width=1.31\columnwidth]{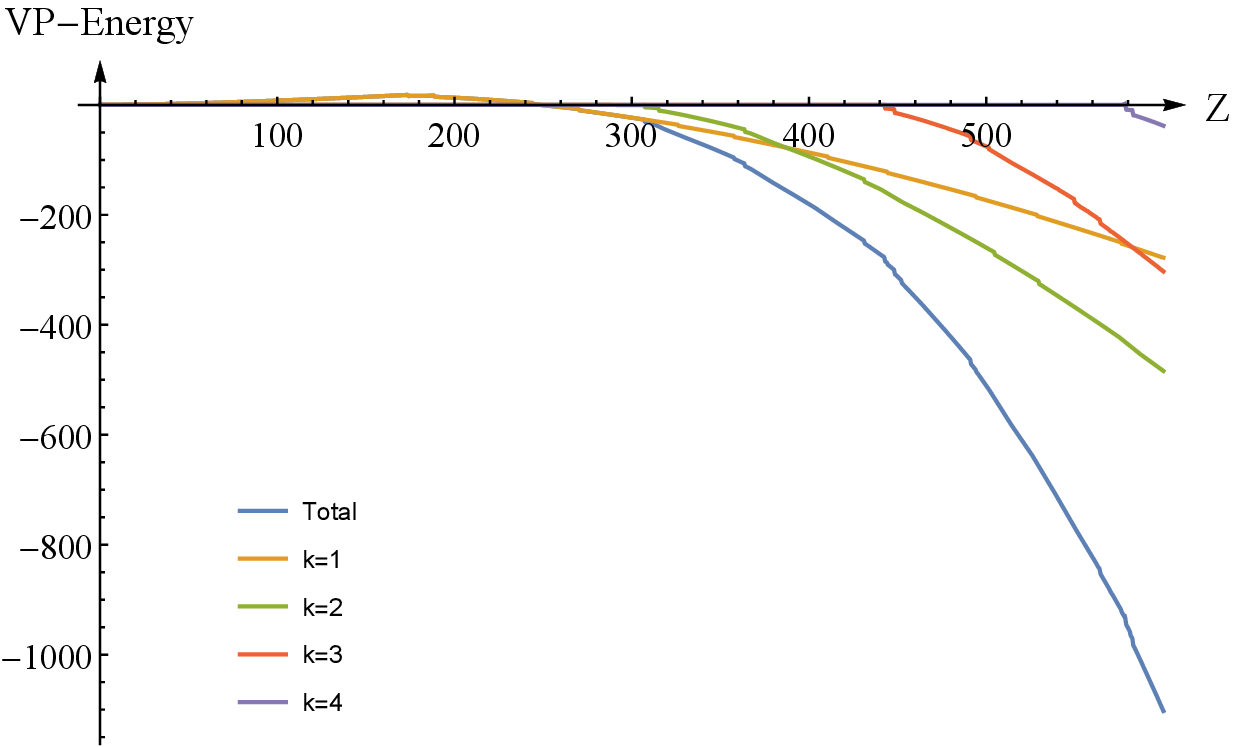}
}
\caption{(Color online) $I(k)\, , \S(k)\, , \E_{VP,k}^{ren}(Z)$ and  $I_{tot}=\sum_{k=1}^{4}I(k)\, , S_{tot}=\sum_{k=1}^{4}S(k)\, , \E_{VP,tot}^{ren}(Z)=\sum_{k=1}^{4}\E_{VP,k}^{ren}(Z) $ on the interval  $0 < Z < 600$ . }
	\label{Int-Sum-VP-600}	
\end{figure*}

First, in Fig.\ref{Int-Sum-VP-600} there are shown the specific features of partial  phase integrals
\beq
\label{3.42}
I(k)={k \over \pi}\, \int\limits_0^{\inf} \!   \  \frac{q \, d q }{\sqrt{q^2+1}} \ \d_{tot}(k,q) \ ,
 \eeq
partial sums over discrete levels
\beq
\label{5.15}
S(k)=k\,\sum\limits_{\pm} \sum\limits_{-1 \leqslant \e_{n,\pm k}<1} \(1-\e_{n,\pm k}\) \ ,
\eeq
and renormalized partial VP-energies (\ref{3.59}).

 As it follows from  Fig.\ref{Int-Sum-VP-600}(a),  phase integrals increase monotonically with growing  $ Z $ and are always positive. The clearly seen  bending, which is in fact nothing else but the negative jump in the derivative of the curve, for each channel  starts  when the first discrete level from this channel attains the lower continuum.   The origin of this effect  is the sharp jump by  $\pi$ in $\d_{tot}(k,q)$ due to the resonance just born. In each $k$-channel such effect is most pronounced for the first level diving, which takes place at  $Z_{cr,1}\simeq 173.6$ for $k=1$, at  $Z_{cr,5}\simeq 307.4$ for $k=2$, at  $Z_{cr,15}\simeq 442.7$ for k=3 and at $Z_{cr,26}\simeq 578.6$ for k=4 \footnote{The critical charges are found in this case by means of the procedure outlined in Ref.~\cite{Greiner2012}.}.

 The subsequent levels diving also leads to jumps in the derivative of $I(k)$, but they turn out to be much less pronounced already, since with increasing $Z_{cr}$  just below the threshold of the lower continuum for $Z=Z_{cr}+\D Z \ , \ \D Z \ll Z_{cr}$, the  resonance broadening and its rate of further diving into the lower continuum are exponentially slower in agreement  with the well-known result~\cite{Zeldovich1972}, according to which the resonance width just under the threshold behaves like $\sim \exp \( - \sqrt{Z_{cr}/\D Z}\)$. This effect leads to that for each subsequent resonance the region of the phase jump by $\pi$ with increasing  $Z$ grows exponentially slower, and derivative of the phase integral changes in the same way. At the same time, before first levels diving the curves of $I(k)$ in all the partial  channels show up  an almost quadratical growth. For the last channel with $k=4$ this growth takes place during almost the whole interval $0 < Z < 600$, since the first level diving in this channel occurs at $Z \simeq 578.6$.

The  behavior of the total bound energy of discrete levels per each partial channel $S(k)$  is shown in Fig.\ref{Int-Sum-VP-600}(b). $S(k)$  are  discontinuous functions with jumps emerging each time, when the charge of the source reaches the subsequent critical value and the corresponding discrete level dives into the lower continuum. At this moment  the bound energy loses exactly two units of $mc^2 $, which in the final answer must be multiplied by the degeneracy factor $k$. Due to this factor the jumps in the curves of $S(k)$ are more pronounced with growing $k$. On the intervals between two neighboring  $ Z_{cr} $ the bound energy is always positive and increases monotonically, since there grow the bound energies of all the  discrete levels, while on the intervals between $Z=0$ up to first levels diving their growth is almost quadratic.

The partial VP-energies $ \E_{VP,k}^{ren}(Z) $ are shown in Fig.\ref{Int-Sum-VP-600}(c). Note that the behavior of $k=1$ channel is different from the others, since in this channel the structure of the renormalization coefficient  $ \z_{1} $ differs from those with $k>1$ by the   perturbative (Uehling) contribution to VP-energy $\E^{(1)}_{VP}(Z)$.
It is indeed the latter term in the total VP-energy, which is responsible for an almost quadratic growth with $Z$ of VP-energy on the interval $0 < Z < Z_{cr,1}$.
The global change in the behavior of $ \E_{VP,1}^{ren}(Z) $ from the perturbative quadratic growth for $Z \ll Z_{cr,1}$, when the dominant contribution comes from  $\E^{(1)}_{VP}(Z)$, to the regime of decrease into the negative range with increasing $Z$ beyond  $Z_{cr,1}$ is shown  below in Figs.\ref{VP240},\ref{VPball240}.
\begin{figure*}[ht!]
\subfigure[]{
		\includegraphics[width=\columnwidth]{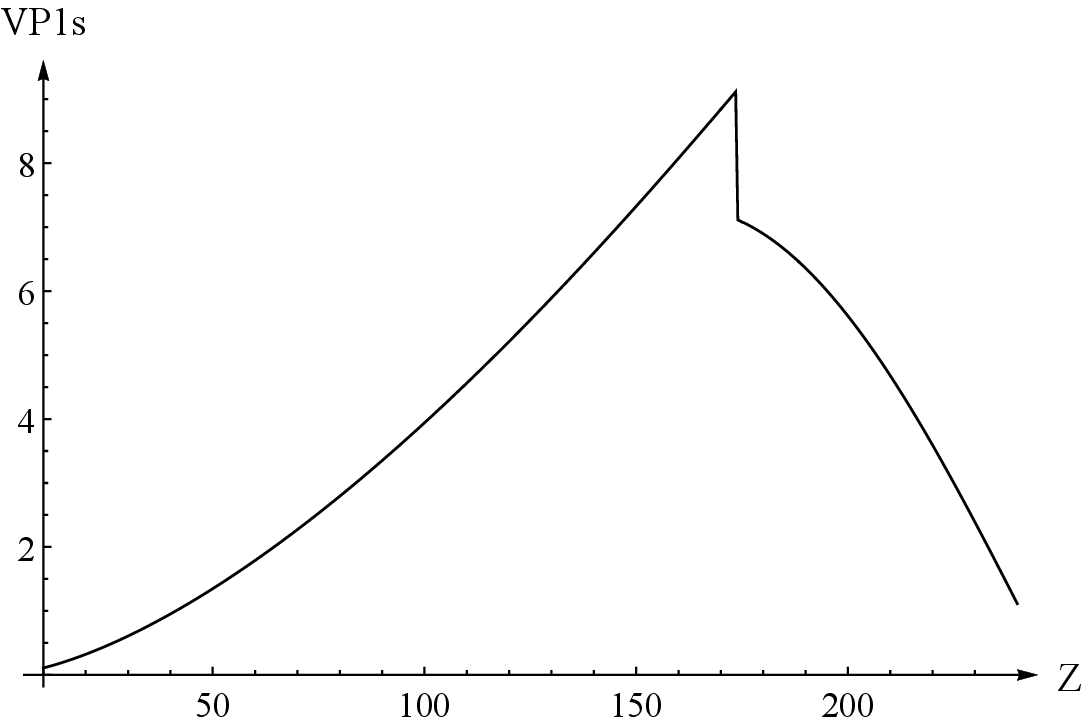}
}
\hfill
\subfigure[]{
		\includegraphics[width=\columnwidth]{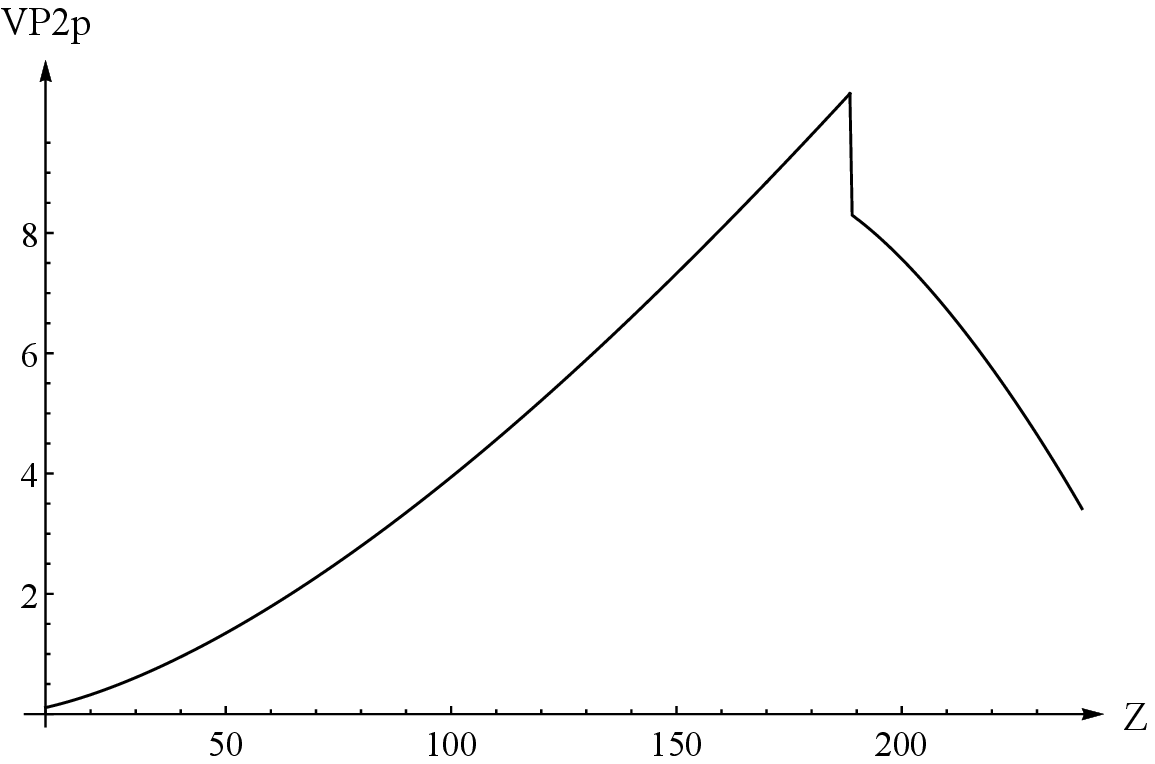}
}
\vfill
\subfigure[]{
		\includegraphics[width=1.2\columnwidth]{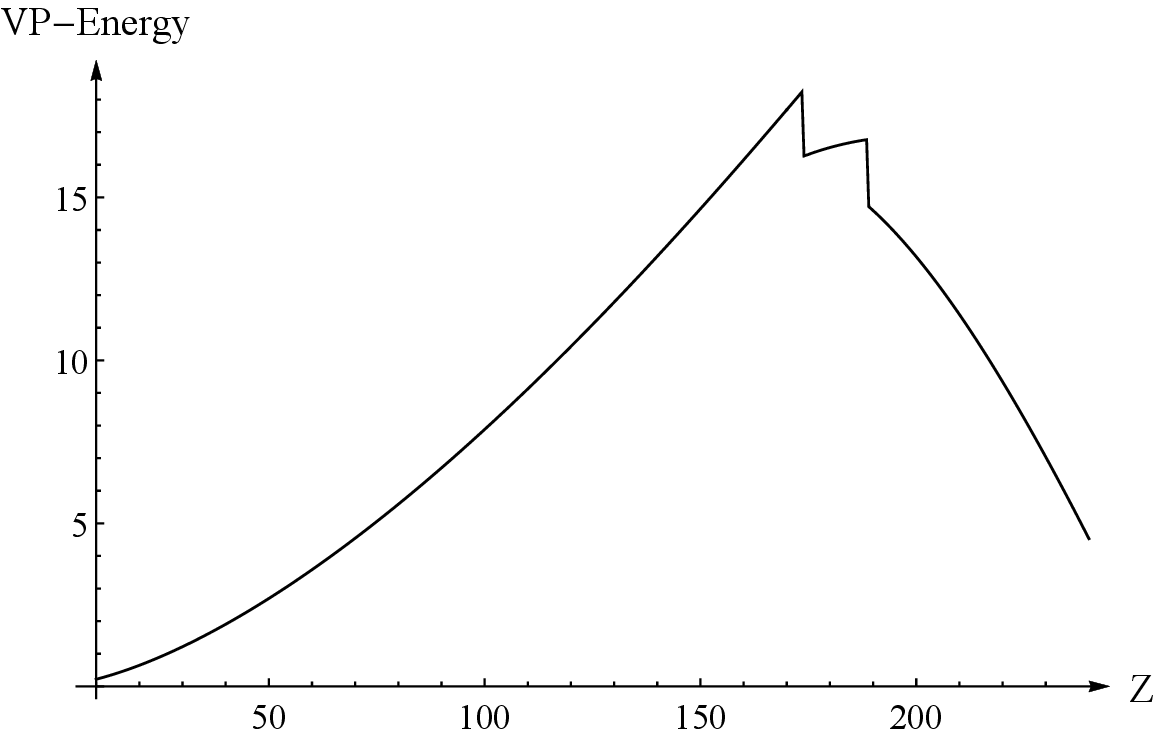}
}
\caption{$ \E_{VP,+}^{ren}(Z)\,, \ \E_{VP,-}^{ren}(Z)\,, \ \E_{VP}^{ren}(Z)  $  on the interval  $10 \leqslant Z \leqslant 240$ (sphere).  }
	\label{VP240}	
\end{figure*}
\begin{figure*}[ht!]
\subfigure[]{
		\includegraphics[width=\columnwidth]{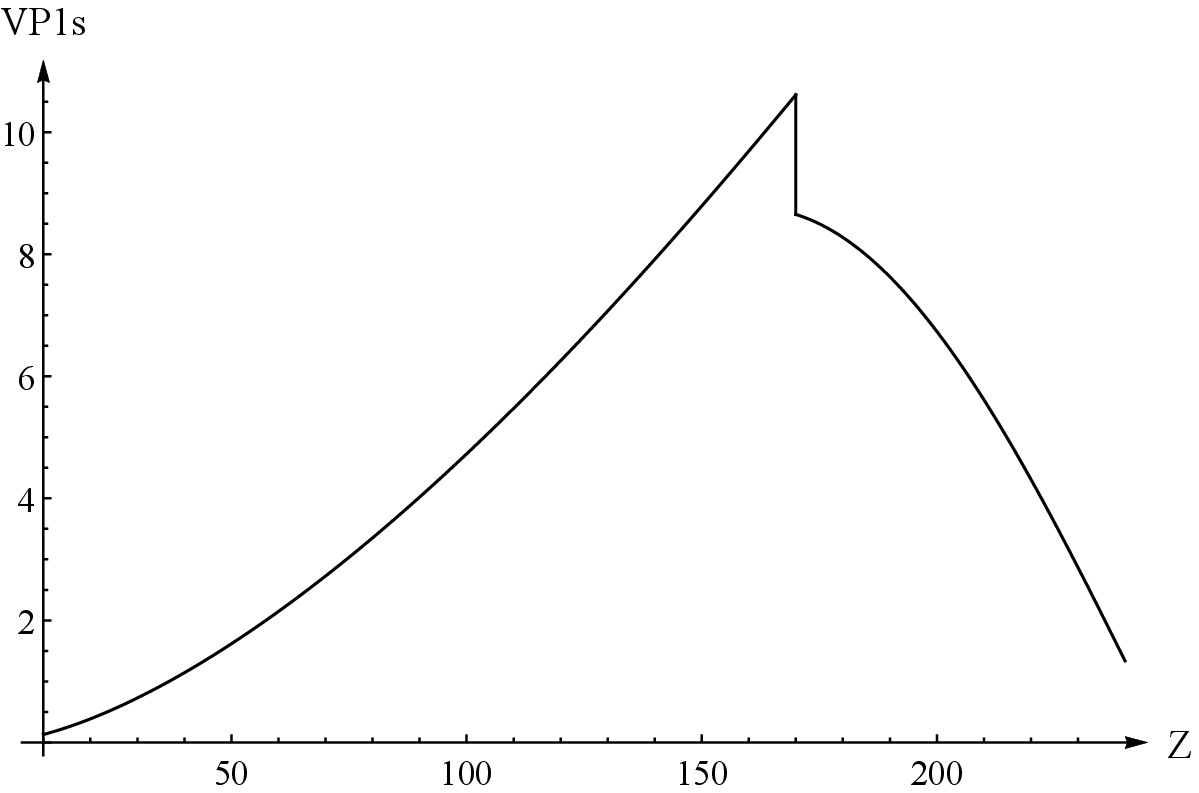}
}
\hfill
\subfigure[]{
		\includegraphics[width=\columnwidth]{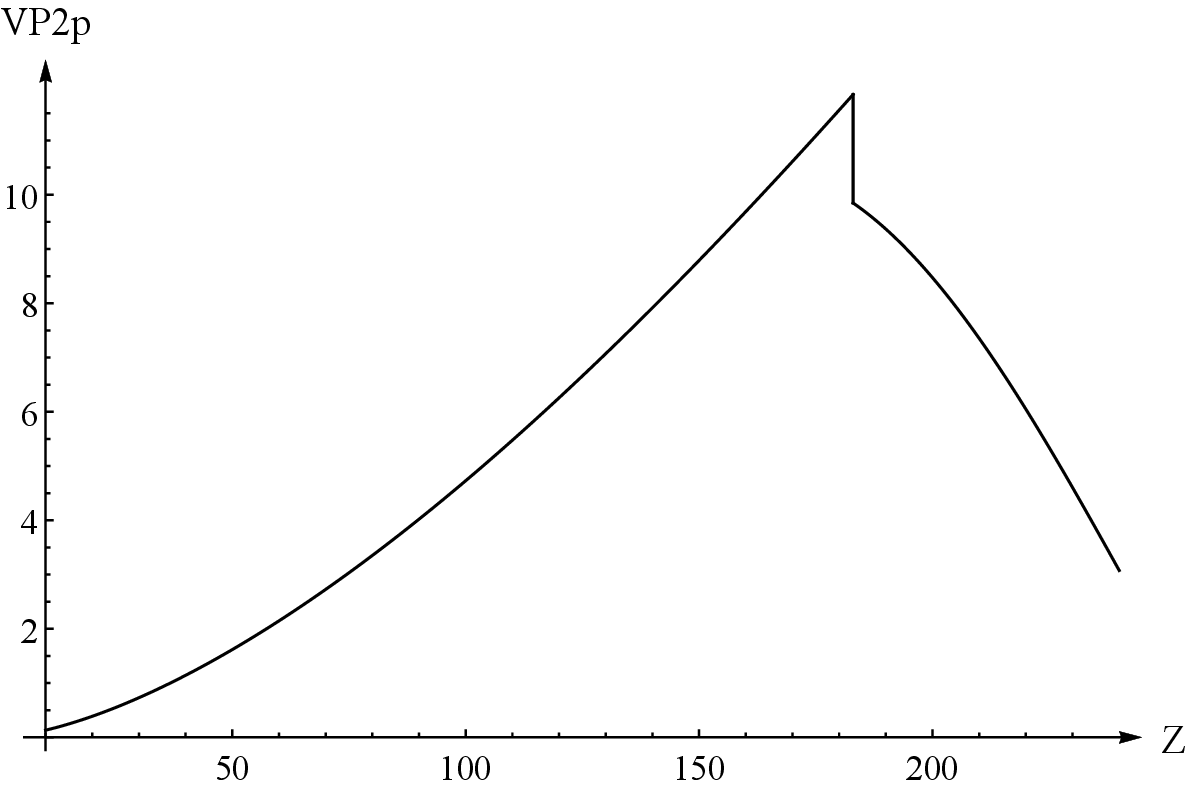}
}
\caption{$ \E_{VP,+}^{ren}(Z)\,, \ \E_{VP,-}^{ren}(Z)\,$  on the interval  $10 \leqslant Z \leqslant 240$ (ball).}
	\label{VPball240}	
\end{figure*}
At the same time, in the  channels  with $k>1$ the quadratic perturbative contribution  is absent. Therefore upon renormalization  (\ref{3.58}-\ref{3.60}), which removes the quadratic component in $I(k)$ and $\S(k)$, just after first level diving $\E_{VP,k}^{ren}(Z) $ reveal with increasing $Z$ a well-pronounced decrease into the negative range.

The final answers for the total VP-energy achieved this way are
\beq
\label{6.2}
\E_{VP}^{ren}(300)=- 23.3 \ , \quad \E_{VP}^{ren}(600)=-1102.7 \ .
\eeq
On this interval of $Z$ the decrease of the total VP-energy into the negative range proceeds very fast,  but with further growth of $Z$ the decay rate becomes smaller. In particular,
\beq
\label{6.3}
\E_{VP}^{ren}(1000)=- 8398.6 \ .
\eeq
while the reasonable estimate of asymptotical behavior of $\E_{VP}^{ren}(Z)$ as a function of $Z$, achieved from the interval $1000 < Z < 3000$, reads
\beq
\label{6.4}
\E_{VP}^{ren}(Z) \sim - Z^4/R(Z) \ .
\eeq

\section{Spontaneous positron emission}

Now --- having dealt with the general properties of VP-effects  this way --- let us consider more thoroughly  the interval $10 \leqslant Z \leqslant 240$, when only two first levels $1s_{1/2}\,, \ 2p_{1/2}$ with opposite parity $(\pm)$ have already dived into the lower continuum at $Z_{cr,1}$ and $Z_{cr,2}$. The behavior of the fixed parity $\E_{VP,\pm}^{ren}(Z)$ for the charged sphere and ball source configurations is shown in Figs.\ref{VP240},\ref{VPball240}.  Note that the only difference between sphere and ball is the small shift of $Z_{cr,i}$ and the general magnitude of VP-energy, which  in the ball case is $\simeq (6/5) \times$VP-energy for the sphere, that is quite close to the ratio of their classical electrostatic self-energies $ 3 Z^2 \a/5 R$ and $ Z^2 \a/2 R$.
\begin{figure*}[ht!]
\subfigure[]{
		\includegraphics[width=\columnwidth]{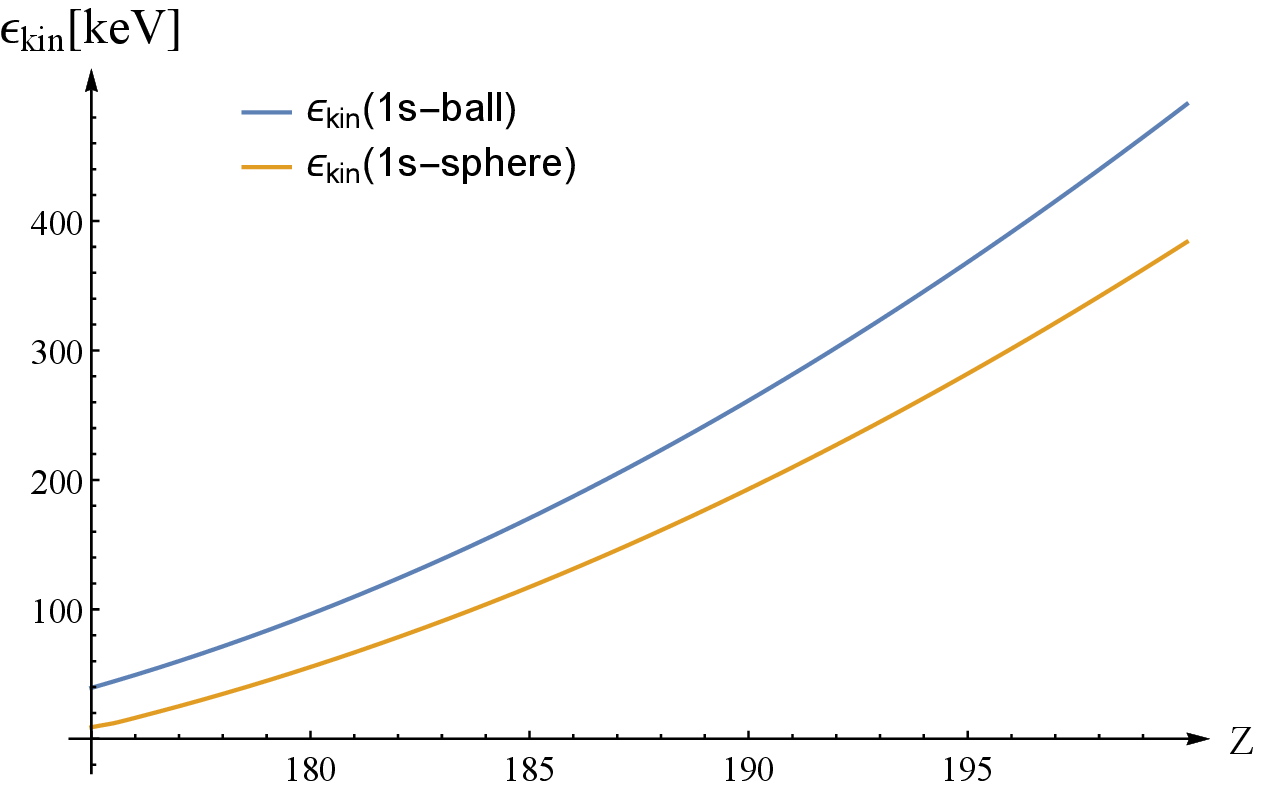}
}
\hfill
\subfigure[]{
		\includegraphics[width=\columnwidth]{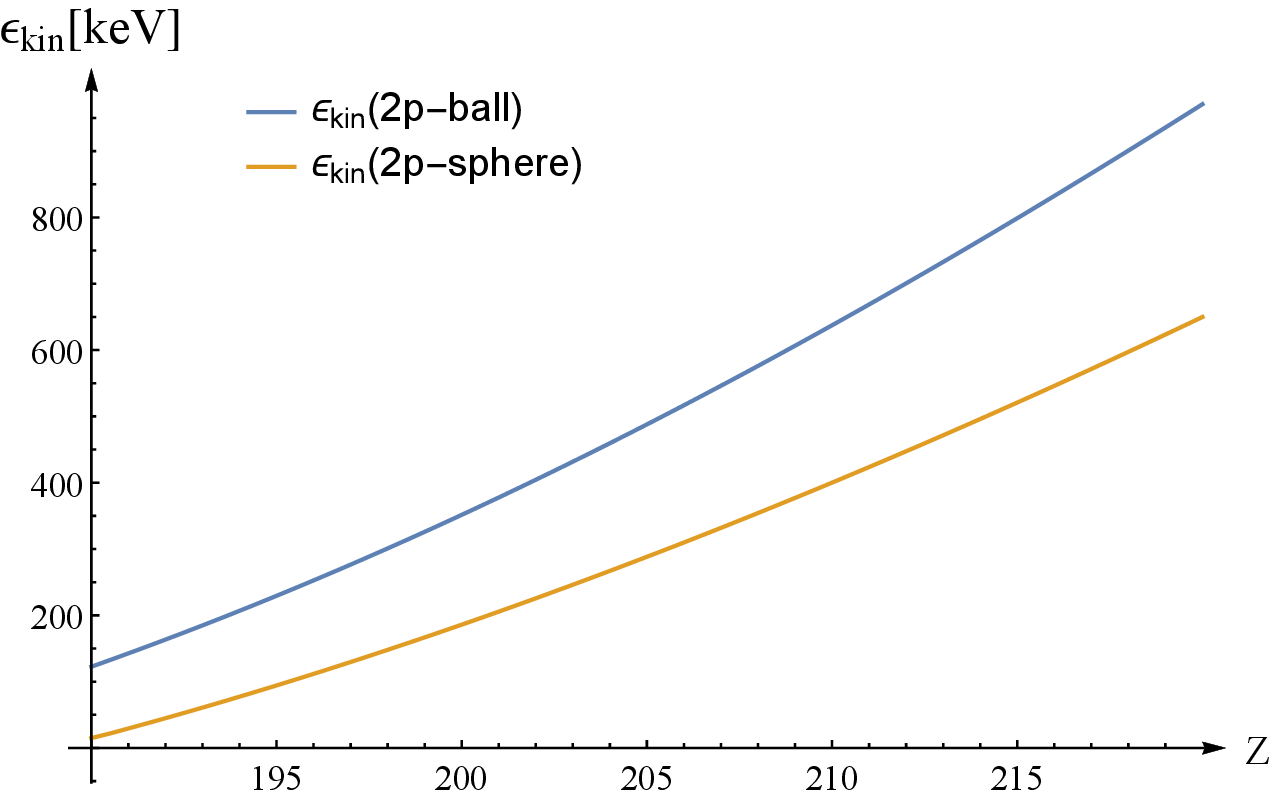}
}
\caption{(Color online)\, $ \e_{kin}(Z)$ with definite parity on the pertinent intervals  of $Z$ for sphere and  ball source configurations and (a): 1s-channel (even); (b): 2p-channel (odd).}
	\label{ekin}	
\end{figure*}
\begin{figure*}[ht!]
\subfigure[]{
		\includegraphics[width=1.4\columnwidth]{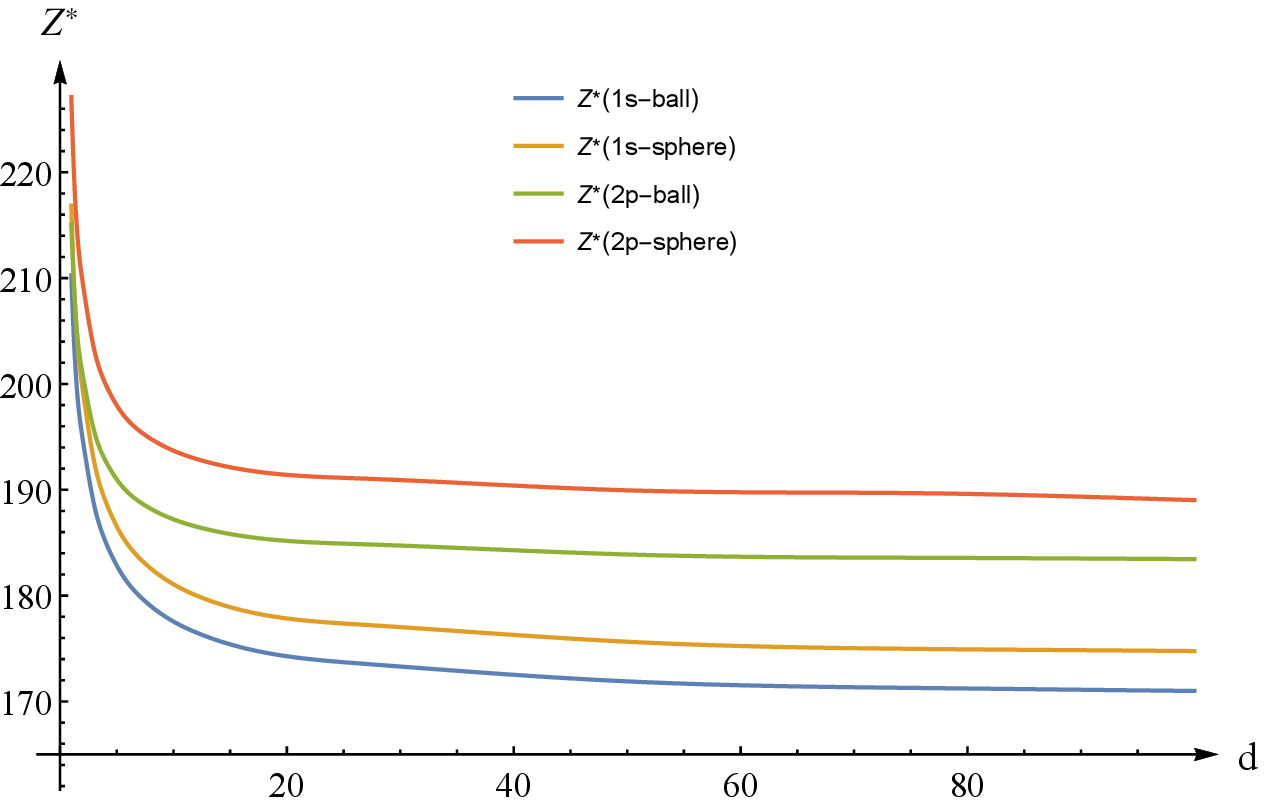}
}
\caption{(Color online)\, $Z^{\ast}(d)$  for  $1\leqslant d \leqslant 100$. }
	\label{Zast-ball-sphere}	
\end{figure*}

General theory~\cite{Greiner1985a,Plunien1986,Greiner2012, Rafelski2016}, based on the  framework~\cite{Fano1961}, predicts that  after diving such a level transforms into a metastable state with lifetime $\sim 10^{-19}$ sec, afterwards there occurs the spontaneous positron emission accompanied with vacuum shells formation  according to  Fano rule (\ref{3.28}).  An important point here is  that due to spherical symmetry of the source during this process all the angular quantum numbers and parity of the dived level are preserved by metastable state and further by positrons created. Furthermore, the spontaneous emission of positrons should be caused solely by VP-effects  without any other channels of energy transfer.

So for each parity  the energy balance suggests the following picture of this process. The rest mass of positrons is created just after level diving via negative jumps in the VP-energy at corresponding $Z_{cr,i}$, which are exactly equal to $2 \times mc^2$ in accordance with two possible spin projections. However,  to create a real positron scattering state, which provides  a sufficiently large  probability of being in the immediate vicinity of the Coulomb source, it is not enough due to electrostatic reflection  between positron and Coulomb source. This circumstance has been outlined first in Refs. ~\cite{Reinhardt1981,*Mueller1988,Ackad2008}, and explored quite recently with more details in Refs.~\cite{Popov2018,*Novak2018,*Maltsev2018,
Maltsev2019,*Maltsev2020}. From the analysis presented above there follows that to supply the emerging vacuum positrons with corresponding reflection energy, an additional decrease of  $\E_{VP}^{ren}(Z)$ in each parity channel is required. So we are led to the following energy balance conditions for spontaneous positrons emission after level diving at $Z_{cr,i}$
\beq
\label{6.5}
\E_{VP,\pm}^{ren}(Z_{cr,i}+0)  - \E_{VP,\pm}^{ren}(Z) =2\,\e_{kin}(Z)  \ ,
\eeq
where $\e_{kin}(Z)$ is the positron kinetic energy and it is assumed that for each parity positrons are created in pairs with opposite spin projections. In agreement with Refs.~\cite{Popov2018,*Novak2018,*Maltsev2018,
Maltsev2019,*Maltsev2020}, in the considered range of $Z$ the spontaneous positrons are limited to the energy $0< \e_{kin}(Z) < 800$ (see Fig.\ref{ekin}), while the related natural resonance widths do not exceed a few KeV~\cite{Marsman2011,*Maltsev2020a}\footnote{In the spherically-symmetric case the most direct way to find the widths of resonances is provided by $\d_{tot}(k,q)$, introduced in eq. (\ref{3.31}). For details see~\cite{Sveshnikov2022}.}.

 It is also useful to introduce the parameter $d$ by means of relation
\beq
\label{6.7}
\e_{kin}(Z)= Z \a/d \ ,
\eeq
which can be interpreted as the distance from the center of the Coulomb source and the conditional point of the vacuum positron creation\footnote{Due to uncertainty relation there isn't any definite point of the vacuum positron creation in the scattering state with fixed energy. However, treatment of the parameter $d$ via distance between conditional point of the vacuum positron creation and the Coulomb  center turns out to be quite pertinent  and besides, can be reliably justified at least in  the quasiclassical approximation.}.
Upon solving   eqs. (\ref{6.5},\ref{6.7}) with respect to $Z$, one finds that the vacuum positron emission  is quite sensitive to  $d$, as it is clearly seen in Fig.\ref{Zast-ball-sphere}. The  reasonable choice for  $d$  is approximately one electron Compton length $\l_C$ ($=1$ in the units accepted). The latter is a rough estimate for the average radius of vacuum shells, created simultaneously with the positron emission.  Therefore it provides the most favorable conditions for  positron production, from the point of view of both the charge distribution and the creation of a specific  lepton pair.

However, the calculations performed show unambiguously that for such $d$ the emission of vacuum positrons  cannot occur earlier than $Z$ exceeds $Z^{\ast}(1s)=217\,, Z^{\ast}(2p)=227$ (sphere), $Z^{\ast}(1s)=211\, , Z^{\ast}(2p)=215$ (ball) (to compare with  $Z_{cr,1}=173.6\,, Z_{cr,2}=188.5$ (sphere), $Z_{cr,1}=170\,, Z_{cr,2}=183.1$ (ball)). Note also that $Z^{\ast}$ increase very rapidly for $d < \l_C$. In particular, already for  $d=2/3$ one obtains $Z^{\ast}(1s)=233\, , Z^{\ast}(2p)=244$ (sphere), $Z^{\ast}(1s)=224\, , Z^{\ast}(2p)=229$ (ball). Such $Z^{\ast}$ lie far beyond  the interval $170 \leqslant Z \leqslant 192 $, which is nowadays the main region of theoretical and experimental activity in heavy ions collisions aimed at the study of such VP-effects ~\cite{Popov2018,Maltsev2019,FAIR2009,Ter2015,MA2017169}. In the ball source configuration the condition $170 \leqslant Z^{\ast} \leqslant 192 $ is fulfilled  by $1s$-channel  for  $d > 3 \l_C$ and  $d > 5 \l_C$ by $2p$-channel, but although such result can considered as the encouraging one, it should be noted that it is achieved within the model (\ref{1.6},\ref{1.8}) with a strongly underestimated  size of the Coulomb source in comparison with the real two heavy-ion super-critical configuration.

It would be worth to note that if allowed by the lepton number, spontaneous positrons can  by no means be emitted also just beyond the corresponding diving point. In particular,  for $d=137$ (one Bohr radius) one obtains  $Z^{\ast}(1s) \simeq 174.0\, , Z^{\ast}(2p) \simeq 188.8$ (sphere),\,$Z^{\ast}(1s) \simeq 170.7\, , Z^{\ast}(2p) \simeq 183.3$ (ball), what is already quite close to corresponding $Z_{cr,i}$. But in this case they appear in states localized far enough from the Coulomb source with small $\e_{kin} \sim 0.01$ and so cannot be distinguished from those emerging due to nuclear conversion.  To the contrary, vacuum positrons created with  $d \simeq \l_C$  possess a number of specific properties~\cite{Popov2018,*Novak2018,*Maltsev2018,
Maltsev2019,*Maltsev2020,Greiner1985a}, which allow for an unambiguous detection, but the charge of the Coulomb source should be taken in this case not less than  $Z^{\ast} \simeq 210$. The negative result of early investigations at GSI~\cite{Mueller1994} can be  at least partially explained by the last circumstance. Moreover, estimating the self-energy contribution to the radiative part of QED effects due to virtual photon exchange near the lower continuum  shows~\cite{Roenko2018} that it is just a perturbative correction to essentially nonlinear VP-effects caused by  fermionic loop and so yields only a small increase of  $Z_{cr,i}$ and of other VP-quantities under question.

\section{Concluding remarks}

Thus,  lepton number poses serious questions for both theory and experiment dealing with Coulomb super-criticality. Any reliable answer concerning the spontaneous positron emission --- either positive or negative --- is important for our understanding of the nature of this number. For the quantities like charge and spin, which should be also shared by vacuum shells if the spontaneous emission takes place, there are much less questions due to extended models of baryons as topological quantum solitons, including skyrmeons~\cite{ADKINS1983552,*Holzwarth_1986}, chiral quark-pion models~\cite{Weigel2007} and/or chiral bags~\cite{Hosaka2001}. Apart from charge and spin, which can be defined as extended quantities for a quantum soliton in a general framework~\cite{Rajaraman1982,DUBIKOVSKY199480}, these models define also the baryon number by means of the non-trivial topology of underlying $\s$-fields as a conserved quantity with extended spatial distribution. But the lepton number is different, since so far leptons show up as point-like particles with no indications on existence of any kind intrinsic structure.

Recent papers~\cite{Popov2018,*Novak2018,*Maltsev2018,
Maltsev2019,*Maltsev2020}, aimed at the detailed study of spontaneous positron emission in slow heavy-ion collisions, have shown that one could expect a clear signal of transition to the super-critical mode for  bare nuclei with the highest $Z=96$, whose colliding trajectories are close to head-on ones. These results look quite promising, but one should keep in mind that the slow head-on collisions of charged particles are highly unstable with respect to deviations in the transverse plane. Meanwhile, the case under consideration implies the scenario of colliding beams with total number of particles not less than $10^6$, hence the deviations of colliding trajectories due to interparticle interactions in the beam are inevitable. The large mass of nuclei  does not matter, since it is compensated by the almost equally large charge.  So one should expect that the most part of such slow collisions reduces to the peripheral ones, which cannot be reliably distinguished from the nuclear conversion.  To reduce such instability one should consider more complicated collisions, which are based on the regular polyhedron symmetry~\cite{Weyl2015} --- synchronized slow heavy-ion beams move  from the vertices of the polyhedron towards its center. The simplest collision of such kind contains 4 beams, arranged as the mean lines of the tetrahedron with the intersection angle $\vt=2 \arcsin (\sqrt{2/3}) \simeq 109^{\small 0}\,28'$, akin to s-p hybridized tetrahedron carbon bonds. The next and probably the most preferable one is the 6-beam configuration, reproducing  6 semi-axes of the rectangular coordinate frame \footnote{Because the other Platon polygons contain already 8, 12 and 20 vertices,  it seems doubtful that such highly symmetric systems of synchronized beams  can be realized by existing experimental facilities.}. The advantage of such many-beam collisions is also that it requires significantly lower ion charges to attain the super-critical region. On the other hand, such collisions require serious additional efforts for their implementation. Therefore, an additional tasty candy is needed to arouse sufficient interest in such a project, and it is really possible, but this issue will be discussed in a separate paper.

\section{Acknowledgments}

The authors are very indebted to Dr. O.V.Pavlovsky and P.A.Grashin from MSU Department of Physics and to A.S.Davydov from Kurchatov Center for interest and helpful discussions.  This work has been supported in part by the RF Ministry of Sc. $\&$ Ed.  Scientific Research Program, projects No. 01-2014-63889, A16-116021760047-5, and by RFBR grant No. 14-02-01261. The research is carried out using the equipment of the shared research facilities of HPC computing resources at Lomonosov Moscow State University.

\bibliography{VP3DC}

\begin{thebibliography}{51}%
\makeatletter
\providecommand \@ifxundefined [1]{%
 \@ifx{#1\undefined}
}%
\providecommand \@ifnum [1]{%
 \ifnum #1\expandafter \@firstoftwo
 \else \expandafter \@secondoftwo
 \fi
}%
\providecommand \@ifx [1]{%
 \ifx #1\expandafter \@firstoftwo
 \else \expandafter \@secondoftwo
 \fi
}%
\providecommand \natexlab [1]{#1}%
\providecommand \enquote  [1]{``#1''}%
\providecommand \bibnamefont  [1]{#1}%
\providecommand \bibfnamefont [1]{#1}%
\providecommand \citenamefont [1]{#1}%
\providecommand \href@noop [0]{\@secondoftwo}%
\providecommand \href [0]{\begingroup \@sanitize@url \@href}%
\providecommand \@href[1]{\@@startlink{#1}\@@href}%
\providecommand \@@href[1]{\endgroup#1\@@endlink}%
\providecommand \@sanitize@url [0]{\catcode `\\12\catcode `\$12\catcode
  `\&12\catcode `\#12\catcode `\^12\catcode `\_12\catcode `\%12\relax}%
\providecommand \@@startlink[1]{}%
\providecommand \@@endlink[0]{}%
\providecommand \url  [0]{\begingroup\@sanitize@url \@url }%
\providecommand \@url [1]{\endgroup\@href {#1}{\urlprefix }}%
\providecommand \urlprefix  [0]{URL }%
\providecommand \Eprint [0]{\href }%
\providecommand \doibase [0]{http://dx.doi.org/}%
\providecommand \selectlanguage [0]{\@gobble}%
\providecommand \bibinfo  [0]{\@secondoftwo}%
\providecommand \bibfield  [0]{\@secondoftwo}%
\providecommand \translation [1]{[#1]}%
\providecommand \BibitemOpen [0]{}%
\providecommand \bibitemStop [0]{}%
\providecommand \bibitemNoStop [0]{.\EOS\space}%
\providecommand \EOS [0]{\spacefactor3000\relax}%
\providecommand \BibitemShut  [1]{\csname bibitem#1\endcsname}%
\let\auto@bib@innerbib\@empty
\bibitem [{\citenamefont {Rafelski}\ \emph {et~al.}(2017)\citenamefont
  {Rafelski}, \citenamefont {Kirsch}, \citenamefont {M\"uller}, \citenamefont
  {Reinhardt},\ and\ \citenamefont {Greiner}}]{Rafelski2016}%
  \BibitemOpen
  \bibfield  {author} {\bibinfo {author} {\bibfnamefont {J.}~\bibnamefont
  {Rafelski}}, \bibinfo {author} {\bibfnamefont {J.}~\bibnamefont {Kirsch}},
  \bibinfo {author} {\bibfnamefont {B.}~\bibnamefont {M\"uller}}, \bibinfo
  {author} {\bibfnamefont {J.}~\bibnamefont {Reinhardt}}, \ and\ \bibinfo
  {author} {\bibfnamefont {W.}~\bibnamefont {Greiner}},\ }\enquote {\bibinfo
  {title} {Probing {QED} {Vacuum} with {Heavy} {Ions}},}\ in\ \href {\doibase
  10.1007/978-3-319-44165-8_17} {\emph {\bibinfo {booktitle} {New {Horizons} in
  {Fundamental} {Physics}}}},\ \bibinfo {series and number} {{FIAS}
  {Interdisciplinary} {Science} {Series}}\ (\bibinfo  {publisher} {Springer},\
  \bibinfo {year} {2017})\ pp.\ \bibinfo {pages} {211--251}\BibitemShut
  {NoStop}%
\bibitem [{\citenamefont {Kuleshov}\ \emph
  {et~al.}(2015{\natexlab{a}})\citenamefont {Kuleshov}, \citenamefont {Mur},
  \citenamefont {Narozhny}, \citenamefont {Fedotov},\ and\ \citenamefont
  {Lozovik}}]{Kuleshov2015a}%
  \BibitemOpen
  \bibfield  {author} {\bibinfo {author} {\bibfnamefont {V.~M.}\ \bibnamefont
  {Kuleshov}}, \bibinfo {author} {\bibfnamefont {V.~D.}\ \bibnamefont {Mur}},
  \bibinfo {author} {\bibfnamefont {N.~B.}\ \bibnamefont {Narozhny}}, \bibinfo
  {author} {\bibfnamefont {A.~M.}\ \bibnamefont {Fedotov}}, \ and\ \bibinfo
  {author} {\bibfnamefont {Y.~E.}\ \bibnamefont {Lozovik}},\ }\href {\doibase
  10.1134/s0021364015040098} {\bibfield  {journal} {\bibinfo  {journal} {Jetp
  Lett.}\ }\textbf {\bibinfo {volume} {101}},\ \bibinfo {pages} {264} (\bibinfo
  {year} {2015}{\natexlab{a}})}\BibitemShut {NoStop}%
\bibitem [{\citenamefont {Kuleshov}\ \emph
  {et~al.}(2015{\natexlab{b}})\citenamefont {Kuleshov}, \citenamefont {Mur},
  \citenamefont {Narozhny}, \citenamefont {Fedotov}, \citenamefont {Lozovik},\
  and\ \citenamefont {Popov}}]{Kuleshov2015b}%
  \BibitemOpen
  \bibfield  {author} {\bibinfo {author} {\bibfnamefont {V.~M.}\ \bibnamefont
  {Kuleshov}}, \bibinfo {author} {\bibfnamefont {V.~D.}\ \bibnamefont {Mur}},
  \bibinfo {author} {\bibfnamefont {N.~B.}\ \bibnamefont {Narozhny}}, \bibinfo
  {author} {\bibfnamefont {A.~M.}\ \bibnamefont {Fedotov}}, \bibinfo {author}
  {\bibfnamefont {Y.~E.}\ \bibnamefont {Lozovik}}, \ and\ \bibinfo {author}
  {\bibfnamefont {V.~S.}\ \bibnamefont {Popov}},\ }\href {\doibase
  10.3367/ufne.0185.201508d.0845} {\bibfield  {journal} {\bibinfo  {journal}
  {Physics-Uspekhi}\ }\textbf {\bibinfo {volume} {58}},\ \bibinfo {pages} {785}
  (\bibinfo {year} {2015}{\natexlab{b}})}\BibitemShut {NoStop}%
\bibitem [{\citenamefont {Godunov}\ \emph {et~al.}(2017)\citenamefont
  {Godunov}, \citenamefont {Machet},\ and\ \citenamefont
  {Vysotsky}}]{Godunov2017}%
  \BibitemOpen
  \bibfield  {author} {\bibinfo {author} {\bibfnamefont {S.~I.}\ \bibnamefont
  {Godunov}}, \bibinfo {author} {\bibfnamefont {B.}~\bibnamefont {Machet}}, \
  and\ \bibinfo {author} {\bibfnamefont {M.~I.}\ \bibnamefont {Vysotsky}},\
  }\href {\doibase 10.1140/epjc/s10052-017-5325-4} {\bibfield  {journal}
  {\bibinfo  {journal} {Eur. Phys. J. C}\ }\textbf {\bibinfo {volume} {77}},\
  \bibinfo {pages} {77:782} (\bibinfo {year} {2017})}\BibitemShut {NoStop}%
\bibitem [{\citenamefont {Davydov}\ \emph {et~al.}(2017)\citenamefont
  {Davydov}, \citenamefont {Sveshnikov},\ and\ \citenamefont
  {Voronina}}]{Davydov2017}%
  \BibitemOpen
  \bibfield  {author} {\bibinfo {author} {\bibfnamefont {A.}~\bibnamefont
  {Davydov}}, \bibinfo {author} {\bibfnamefont {K.}~\bibnamefont {Sveshnikov}},
  \ and\ \bibinfo {author} {\bibfnamefont {Y.}~\bibnamefont {Voronina}},\
  }\href {\doibase 10.1142/S0217751X17500543} {\bibfield  {journal} {\bibinfo
  {journal} {Int. J. Mod. Phys. A}\ }\textbf {\bibinfo {volume} {32}},\
  \bibinfo {pages} {1750054} (\bibinfo {year} {2017})}\BibitemShut {NoStop}%
\bibitem [{\citenamefont {Voronina}\ \emph
  {et~al.}(2017{\natexlab{a}})\citenamefont {Voronina}, \citenamefont
  {Davydov},\ and\ \citenamefont {Sveshnikov}}]{Sveshnikov2017}%
  \BibitemOpen
  \bibfield  {author} {\bibinfo {author} {\bibfnamefont {Y.}~\bibnamefont
  {Voronina}}, \bibinfo {author} {\bibfnamefont {A.}~\bibnamefont {Davydov}}, \
  and\ \bibinfo {author} {\bibfnamefont {K.}~\bibnamefont {Sveshnikov}},\
  }\href {\doibase 10.1134/S004057791711006X} {\bibfield  {journal} {\bibinfo
  {journal} {Theor. Math. Phys.}\ }\textbf {\bibinfo {volume} {193}},\ \bibinfo
  {pages} {1647} (\bibinfo {year} {2017}{\natexlab{a}})}\BibitemShut {NoStop}%
\bibitem [{\citenamefont {Voronina}\ \emph
  {et~al.}(2017{\natexlab{b}})\citenamefont {Voronina}, \citenamefont
  {Davydov},\ and\ \citenamefont {Sveshnikov}}]{Voronina2017}%
  \BibitemOpen
  \bibfield  {author} {\bibinfo {author} {\bibfnamefont {Y.}~\bibnamefont
  {Voronina}}, \bibinfo {author} {\bibfnamefont {A.}~\bibnamefont {Davydov}}, \
  and\ \bibinfo {author} {\bibfnamefont {K.}~\bibnamefont {Sveshnikov}},\
  }\href {\doibase 10.1134/S1547477117050144} {\bibfield  {journal} {\bibinfo
  {journal} {Phys. Part. Nucl. Lett.}\ }\textbf {\bibinfo {volume} {14}},\
  \bibinfo {pages} {698 } (\bibinfo {year} {2017}{\natexlab{b}})}\BibitemShut
  {NoStop}%
\bibitem [{\citenamefont {Popov}\ \emph {et~al.}(2018)\citenamefont {Popov},
  \citenamefont {Bondarev}, \citenamefont {Kozhedub}, \citenamefont {Maltsev},
  \citenamefont {Shabaev}, \citenamefont {Tupitsyn}, \citenamefont {Ma},
  \citenamefont {Plunien},\ and\ \citenamefont {St{\"o}hlker}}]{Popov2018}%
  \BibitemOpen
  \bibfield  {author} {\bibinfo {author} {\bibfnamefont {R.}~\bibnamefont
  {Popov}}, \bibinfo {author} {\bibfnamefont {A.}~\bibnamefont {Bondarev}},
  \bibinfo {author} {\bibfnamefont {Y.}~\bibnamefont {Kozhedub}}, \bibinfo
  {author} {\bibfnamefont {I.}~\bibnamefont {Maltsev}}, \bibinfo {author}
  {\bibfnamefont {V.}~\bibnamefont {Shabaev}}, \bibinfo {author} {\bibfnamefont
  {I.}~\bibnamefont {Tupitsyn}}, \bibinfo {author} {\bibfnamefont
  {X.}~\bibnamefont {Ma}}, \bibinfo {author} {\bibfnamefont {G.}~\bibnamefont
  {Plunien}}, \ and\ \bibinfo {author} {\bibfnamefont {T.}~\bibnamefont
  {St{\"o}hlker}},\ }\href {\doibase 10.1140/epjd/e2018-90056-4} {\bibfield
  {journal} {\bibinfo  {journal} {Eur. Phys. J. D}\ }\textbf {\bibinfo {volume}
  {72}},\ \bibinfo {pages} {115} (\bibinfo {year} {2018})}\BibitemShut
  {NoStop}%
\bibitem [{\citenamefont {Novak}\ \emph {et~al.}(2018)\citenamefont {Novak},
  \citenamefont {Kholodov}, \citenamefont {Surzhykov}, \citenamefont
  {Artemyev},\ and\ \citenamefont {St{\"o}hlker}}]{Novak2018}%
  \BibitemOpen
  \bibfield  {author} {\bibinfo {author} {\bibfnamefont {O.}~\bibnamefont
  {Novak}}, \bibinfo {author} {\bibfnamefont {R.}~\bibnamefont {Kholodov}},
  \bibinfo {author} {\bibfnamefont {A.}~\bibnamefont {Surzhykov}}, \bibinfo
  {author} {\bibfnamefont {A.~N.}\ \bibnamefont {Artemyev}}, \ and\ \bibinfo
  {author} {\bibfnamefont {T.}~\bibnamefont {St{\"o}hlker}},\ }\href {\doibase
  https://doi.org/10.1103/PhysRevA.97.032518} {\bibfield  {journal} {\bibinfo
  {journal} {Phys. Rev. A}\ }\textbf {\bibinfo {volume} {97}},\ \bibinfo
  {pages} {032518} (\bibinfo {year} {2018})}\BibitemShut {NoStop}%
\bibitem [{\citenamefont {Maltsev}\ \emph {et~al.}(2018)\citenamefont
  {Maltsev}, \citenamefont {Shabaev}, \citenamefont {Popov}, \citenamefont
  {Kozhedub}, \citenamefont {Plunien}, \citenamefont {Ma},\ and\ \citenamefont
  {St{\"o}hlker}}]{Maltsev2018}%
  \BibitemOpen
  \bibfield  {author} {\bibinfo {author} {\bibfnamefont {I.~A.}\ \bibnamefont
  {Maltsev}}, \bibinfo {author} {\bibfnamefont {V.~M.}\ \bibnamefont
  {Shabaev}}, \bibinfo {author} {\bibfnamefont {R.~V.}\ \bibnamefont {Popov}},
  \bibinfo {author} {\bibfnamefont {Y.~S.}\ \bibnamefont {Kozhedub}}, \bibinfo
  {author} {\bibfnamefont {G.}~\bibnamefont {Plunien}}, \bibinfo {author}
  {\bibfnamefont {X.}~\bibnamefont {Ma}}, \ and\ \bibinfo {author}
  {\bibfnamefont {T.}~\bibnamefont {St{\"o}hlker}},\ }\href {\doibase
  10.1103/PhysRevA.98.062709} {\bibfield  {journal} {\bibinfo  {journal} {Phys.
  Rev. A}\ }\textbf {\bibinfo {volume} {98}},\ \bibinfo {pages} {062709}
  (\bibinfo {year} {2018})}\BibitemShut {NoStop}%
\bibitem [{\citenamefont {Roenko}\ and\ \citenamefont
  {Sveshnikov}(2018)}]{Roenko2018}%
  \BibitemOpen
  \bibfield  {author} {\bibinfo {author} {\bibfnamefont {A.}~\bibnamefont
  {Roenko}}\ and\ \bibinfo {author} {\bibfnamefont {K.}~\bibnamefont
  {Sveshnikov}},\ }\href {\doibase https://doi.org/10.1103/PhysRevA.97.012113}
  {\bibfield  {journal} {\bibinfo  {journal} {Phys. Rev. A}\ }\textbf {\bibinfo
  {volume} {97}},\ \bibinfo {pages} {012113} (\bibinfo {year}
  {2018})}\BibitemShut {NoStop}%
\bibitem [{\citenamefont {Maltsev}\ \emph {et~al.}(2019)\citenamefont
  {Maltsev}, \citenamefont {Shabaev}, \citenamefont {Popov}, \citenamefont
  {Kozhedub}, \citenamefont {Plunien}, \citenamefont {Ma}, \citenamefont
  {St\"ohlker},\ and\ \citenamefont {Tumakov}}]{Maltsev2019}%
  \BibitemOpen
  \bibfield  {author} {\bibinfo {author} {\bibfnamefont {I.~A.}\ \bibnamefont
  {Maltsev}}, \bibinfo {author} {\bibfnamefont {V.~M.}\ \bibnamefont
  {Shabaev}}, \bibinfo {author} {\bibfnamefont {R.~V.}\ \bibnamefont {Popov}},
  \bibinfo {author} {\bibfnamefont {Y.~S.}\ \bibnamefont {Kozhedub}}, \bibinfo
  {author} {\bibfnamefont {G.}~\bibnamefont {Plunien}}, \bibinfo {author}
  {\bibfnamefont {X.}~\bibnamefont {Ma}}, \bibinfo {author} {\bibfnamefont
  {T.}~\bibnamefont {St\"ohlker}}, \ and\ \bibinfo {author} {\bibfnamefont
  {D.~A.}\ \bibnamefont {Tumakov}},\ }\href {\doibase
  10.1103/PhysRevLett.123.113401} {\bibfield  {journal} {\bibinfo  {journal}
  {Phys. Rev. Lett.}\ }\textbf {\bibinfo {volume} {123}},\ \bibinfo {pages}
  {113401} (\bibinfo {year} {2019})}\BibitemShut {NoStop}%
\bibitem [{\citenamefont {Popov}\ \emph {et~al.}(2020)\citenamefont {Popov},
  \citenamefont {Shabaev}, \citenamefont {Telnov}, \citenamefont {Tupitsyn},
  \citenamefont {Maltsev}, \citenamefont {Kozhedub}, \citenamefont {Bondarev},
  \citenamefont {Kozin}, \citenamefont {Ma}, \citenamefont {Plunien},
  \citenamefont {St\"ohlker}, \citenamefont {Tumakov},\ and\ \citenamefont
  {Zaytsev}}]{Maltsev2020}%
  \BibitemOpen
  \bibfield  {author} {\bibinfo {author} {\bibfnamefont {R.~V.}\ \bibnamefont
  {Popov}}, \bibinfo {author} {\bibfnamefont {V.~M.}\ \bibnamefont {Shabaev}},
  \bibinfo {author} {\bibfnamefont {D.~A.}\ \bibnamefont {Telnov}}, \bibinfo
  {author} {\bibfnamefont {I.~I.}\ \bibnamefont {Tupitsyn}}, \bibinfo {author}
  {\bibfnamefont {I.~A.}\ \bibnamefont {Maltsev}}, \bibinfo {author}
  {\bibfnamefont {Y.~S.}\ \bibnamefont {Kozhedub}}, \bibinfo {author}
  {\bibfnamefont {A.~I.}\ \bibnamefont {Bondarev}}, \bibinfo {author}
  {\bibfnamefont {N.~V.}\ \bibnamefont {Kozin}}, \bibinfo {author}
  {\bibfnamefont {X.}~\bibnamefont {Ma}}, \bibinfo {author} {\bibfnamefont
  {G.}~\bibnamefont {Plunien}}, \bibinfo {author} {\bibfnamefont
  {T.}~\bibnamefont {St\"ohlker}}, \bibinfo {author} {\bibfnamefont {D.~A.}\
  \bibnamefont {Tumakov}}, \ and\ \bibinfo {author} {\bibfnamefont {V.~A.}\
  \bibnamefont {Zaytsev}},\ }\href {\doibase 10.1103/PhysRevD.102.076005}
  {\bibfield  {journal} {\bibinfo  {journal} {Phys. Rev. D}\ }\textbf {\bibinfo
  {volume} {102}},\ \bibinfo {pages} {076005} (\bibinfo {year}
  {2020})}\BibitemShut {NoStop}%
\bibitem [{\citenamefont {Greiner}\ \emph {et~al.}(1985)\citenamefont
  {Greiner}, \citenamefont {M\"uller},\ and\ \citenamefont
  {Rafelski}}]{Greiner1985a}%
  \BibitemOpen
  \bibfield  {author} {\bibinfo {author} {\bibfnamefont {W.}~\bibnamefont
  {Greiner}}, \bibinfo {author} {\bibfnamefont {B.}~\bibnamefont {M\"uller}}, \
  and\ \bibinfo {author} {\bibfnamefont {J.}~\bibnamefont {Rafelski}},\ }\href
  {http://link.springer.com/book/10.1007/978-3-642-82272-8} {\emph {\bibinfo
  {title} {Quantum Electrodynamics of Strong Fields}}},\ \bibinfo {edition}
  {2nd}\ ed.\ (\bibinfo  {publisher} {Springer},\ \bibinfo {address} {Berlin},\
  \bibinfo {year} {1985})\BibitemShut {NoStop}%
\bibitem [{\citenamefont {Plunien}\ \emph {et~al.}(1986)\citenamefont
  {Plunien}, \citenamefont {M\"uller},\ and\ \citenamefont
  {Greiner}}]{Plunien1986}%
  \BibitemOpen
  \bibfield  {author} {\bibinfo {author} {\bibfnamefont {G.}~\bibnamefont
  {Plunien}}, \bibinfo {author} {\bibfnamefont {B.}~\bibnamefont {M\"uller}}, \
  and\ \bibinfo {author} {\bibfnamefont {W.}~\bibnamefont {Greiner}},\ }\href
  {\doibase 10.1016/0370-1573(86)90020-7} {\bibfield  {journal} {\bibinfo
  {journal} {Phys. Rep.}\ }\textbf {\bibinfo {volume} {134}},\ \bibinfo {pages}
  {87 } (\bibinfo {year} {1986})}\BibitemShut {NoStop}%
\bibitem [{\citenamefont {Greiner}\ and\ \citenamefont
  {Reinhardt}(2009)}]{Greiner2012}%
  \BibitemOpen
  \bibfield  {author} {\bibinfo {author} {\bibfnamefont {W.}~\bibnamefont
  {Greiner}}\ and\ \bibinfo {author} {\bibfnamefont {J.}~\bibnamefont
  {Reinhardt}},\ }\href {\doibase /10.1007/978-3-540-87561-1} {\emph {\bibinfo
  {title} {Quantum Electrodynamics}}},\ \bibinfo {edition} {4th}\ ed.\
  (\bibinfo  {publisher} {Springer-Verlag Berlin Heidelberg},\ \bibinfo {year}
  {2009})\BibitemShut {NoStop}%
\bibitem [{\citenamefont {Ruffini}\ \emph {et~al.}(2010)\citenamefont
  {Ruffini}, \citenamefont {Vereshchagin},\ and\ \citenamefont
  {Xue}}]{Ruffini2010}%
  \BibitemOpen
  \bibfield  {author} {\bibinfo {author} {\bibfnamefont {R.}~\bibnamefont
  {Ruffini}}, \bibinfo {author} {\bibfnamefont {G.}~\bibnamefont
  {Vereshchagin}}, \ and\ \bibinfo {author} {\bibfnamefont {S.-S.}\
  \bibnamefont {Xue}},\ }\href {\doibase 10.1016/j.physrep.2009.10.004}
  {\bibfield  {journal} {\bibinfo  {journal} {Phys. Rep.}\ }\textbf {\bibinfo
  {volume} {487}},\ \bibinfo {pages} {1 } (\bibinfo {year} {2010})}\BibitemShut
  {NoStop}%
\bibitem [{\citenamefont {Reinhardt}\ \emph {et~al.}(1981)\citenamefont
  {Reinhardt}, \citenamefont {M\"uller},\ and\ \citenamefont
  {Greiner}}]{Reinhardt1981}%
  \BibitemOpen
  \bibfield  {author} {\bibinfo {author} {\bibfnamefont {J.}~\bibnamefont
  {Reinhardt}}, \bibinfo {author} {\bibfnamefont {B.}~\bibnamefont {M\"uller}},
  \ and\ \bibinfo {author} {\bibfnamefont {W.}~\bibnamefont {Greiner}},\ }\href
  {\doibase 10.1103/PhysRevA.24.103} {\bibfield  {journal} {\bibinfo  {journal}
  {Phys. Rev. A}\ }\textbf {\bibinfo {volume} {24}},\ \bibinfo {pages} {103}
  (\bibinfo {year} {1981})}\BibitemShut {NoStop}%
\bibitem [{\citenamefont {M{\"u}ller}\ \emph {et~al.}(1988)\citenamefont
  {M{\"u}ller}, \citenamefont {de~Reus}, \citenamefont {Reinhardt},
  \citenamefont {M{\"u}ller}, \citenamefont {Greiner},\ and\ \citenamefont
  {Soff}}]{Mueller1988}%
  \BibitemOpen
  \bibfield  {author} {\bibinfo {author} {\bibfnamefont {U.}~\bibnamefont
  {M{\"u}ller}}, \bibinfo {author} {\bibfnamefont {T.}~\bibnamefont {de~Reus}},
  \bibinfo {author} {\bibfnamefont {J.}~\bibnamefont {Reinhardt}}, \bibinfo
  {author} {\bibfnamefont {B.}~\bibnamefont {M{\"u}ller}}, \bibinfo {author}
  {\bibfnamefont {W.}~\bibnamefont {Greiner}}, \ and\ \bibinfo {author}
  {\bibfnamefont {G.}~\bibnamefont {Soff}},\ }\href {\doibase
  10.1103/PhysRevA.37.1449} {\bibfield  {journal} {\bibinfo  {journal} {Phys.
  Rev. A}\ }\textbf {\bibinfo {volume} {37}},\ \bibinfo {pages} {1449}
  (\bibinfo {year} {1988})}\BibitemShut {NoStop}%
\bibitem [{\citenamefont {Ackad}\ and\ \citenamefont
  {Horbatsch}(2008)}]{Ackad2008}%
  \BibitemOpen
  \bibfield  {author} {\bibinfo {author} {\bibfnamefont {E.}~\bibnamefont
  {Ackad}}\ and\ \bibinfo {author} {\bibfnamefont {M.}~\bibnamefont
  {Horbatsch}},\ }\href {\doibase 10.1103/PhysRevA.78.062711} {\bibfield
  {journal} {\bibinfo  {journal} {Phys. Rev. A}\ }\textbf {\bibinfo {volume}
  {78}},\ \bibinfo {pages} {062711} (\bibinfo {year} {2008})}\BibitemShut
  {NoStop}%
\bibitem [{\citenamefont {Tupitsyn}\ \emph {et~al.}(2010)\citenamefont
  {Tupitsyn}, \citenamefont {Kozhedub}, \citenamefont {Shabaev}, \citenamefont
  {Deyneka}, \citenamefont {Hagmann}, \citenamefont {Kozhuharov}, \citenamefont
  {Plunien},\ and\ \citenamefont {Stohlker}}]{Tupitsyn2010}%
  \BibitemOpen
  \bibfield  {author} {\bibinfo {author} {\bibfnamefont {I.~I.}\ \bibnamefont
  {Tupitsyn}}, \bibinfo {author} {\bibfnamefont {Y.~S.}\ \bibnamefont
  {Kozhedub}}, \bibinfo {author} {\bibfnamefont {V.~M.}\ \bibnamefont
  {Shabaev}}, \bibinfo {author} {\bibfnamefont {G.~B.}\ \bibnamefont
  {Deyneka}}, \bibinfo {author} {\bibfnamefont {S.}~\bibnamefont {Hagmann}},
  \bibinfo {author} {\bibfnamefont {C.}~\bibnamefont {Kozhuharov}}, \bibinfo
  {author} {\bibfnamefont {G.}~\bibnamefont {Plunien}}, \ and\ \bibinfo
  {author} {\bibfnamefont {T.}~\bibnamefont {Stohlker}},\ }\href {\doibase
  10.1103/PhysRevA.82.042701} {\bibfield  {journal} {\bibinfo  {journal} {Phys.
  Rev. A}\ }\textbf {\bibinfo {volume} {82}},\ \bibinfo {pages} {042701}
  (\bibinfo {year} {2010})}\BibitemShut {NoStop}%
\bibitem [{\citenamefont {Wichmann}\ and\ \citenamefont
  {Kroll}(1956)}]{Wichmann1956}%
  \BibitemOpen
  \bibfield  {author} {\bibinfo {author} {\bibfnamefont {E.~H.}\ \bibnamefont
  {Wichmann}}\ and\ \bibinfo {author} {\bibfnamefont {N.~M.}\ \bibnamefont
  {Kroll}},\ }\href {\doibase 10.1103/PhysRev.101.843} {\bibfield  {journal}
  {\bibinfo  {journal} {Phys. Rev.}\ }\textbf {\bibinfo {volume} {101}},\
  \bibinfo {pages} {843} (\bibinfo {year} {1956})}\BibitemShut {NoStop}%
\bibitem [{\citenamefont {Gyulassy}(1975)}]{Gyulassy1975}%
  \BibitemOpen
  \bibfield  {author} {\bibinfo {author} {\bibfnamefont {M.}~\bibnamefont
  {Gyulassy}},\ }\href {\doibase 10.1016/0375-9474(75)90554-0} {\bibfield
  {journal} {\bibinfo  {journal} {Nucl. Phys. A}\ }\textbf {\bibinfo {volume}
  {244}},\ \bibinfo {pages} {497 } (\bibinfo {year} {1975})}\BibitemShut
  {NoStop}%
\bibitem [{\citenamefont {Brown}\ \emph
  {et~al.}(1975{\natexlab{a}})\citenamefont {Brown}, \citenamefont {Cahn},\
  and\ \citenamefont {McLerran}}]{McLerran1975a}%
  \BibitemOpen
  \bibfield  {author} {\bibinfo {author} {\bibfnamefont {L.}~\bibnamefont
  {Brown}}, \bibinfo {author} {\bibfnamefont {R.}~\bibnamefont {Cahn}}, \ and\
  \bibinfo {author} {\bibfnamefont {L.}~\bibnamefont {McLerran}},\ }\href
  {\doibase 10.1103/PhysRevD.12.581} {\bibfield  {journal} {\bibinfo  {journal}
  {Phys. Rev. D}\ }\textbf {\bibinfo {volume} {12}},\ \bibinfo {pages} {581}
  (\bibinfo {year} {1975}{\natexlab{a}})}\BibitemShut {NoStop}%
\bibitem [{\citenamefont {Brown}\ \emph
  {et~al.}(1975{\natexlab{b}})\citenamefont {Brown}, \citenamefont {Cahn},\
  and\ \citenamefont {McLerran}}]{McLerran1975b}%
  \BibitemOpen
  \bibfield  {author} {\bibinfo {author} {\bibfnamefont {L.}~\bibnamefont
  {Brown}}, \bibinfo {author} {\bibfnamefont {R.}~\bibnamefont {Cahn}}, \ and\
  \bibinfo {author} {\bibfnamefont {L.}~\bibnamefont {McLerran}},\ }\href
  {\doibase 10.1103/PhysRevD.12.596} {\bibfield  {journal} {\bibinfo  {journal}
  {Phys. Rev. D}\ }\textbf {\bibinfo {volume} {12}},\ \bibinfo {pages} {596}
  (\bibinfo {year} {1975}{\natexlab{b}})}\BibitemShut {NoStop}%
\bibitem [{\citenamefont {Brown}\ \emph
  {et~al.}(1975{\natexlab{c}})\citenamefont {Brown}, \citenamefont {Cahn},\
  and\ \citenamefont {McLerran}}]{McLerran1975c}%
  \BibitemOpen
  \bibfield  {author} {\bibinfo {author} {\bibfnamefont {L.}~\bibnamefont
  {Brown}}, \bibinfo {author} {\bibfnamefont {R.}~\bibnamefont {Cahn}}, \ and\
  \bibinfo {author} {\bibfnamefont {L.}~\bibnamefont {McLerran}},\ }\href
  {\doibase 10.1103/PhysRevD.12.609} {\bibfield  {journal} {\bibinfo  {journal}
  {Phys. Rev. D}\ }\textbf {\bibinfo {volume} {12}},\ \bibinfo {pages} {609}
  (\bibinfo {year} {1975}{\natexlab{c}})}\BibitemShut {NoStop}%
\bibitem [{\citenamefont {Mohr}\ \emph {et~al.}(1998)\citenamefont {Mohr},
  \citenamefont {Plunien},\ and\ \citenamefont {Soff}}]{Mohr1998}%
  \BibitemOpen
  \bibfield  {author} {\bibinfo {author} {\bibfnamefont {P.~J.}\ \bibnamefont
  {Mohr}}, \bibinfo {author} {\bibfnamefont {G.}~\bibnamefont {Plunien}}, \
  and\ \bibinfo {author} {\bibfnamefont {G.}~\bibnamefont {Soff}},\ }\href
  {\doibase 10.1016/S0370-1573(97)00046-X} {\bibfield  {journal} {\bibinfo
  {journal} {Phys. Rep.}\ }\textbf {\bibinfo {volume} {293}},\ \bibinfo {pages}
  {227 } (\bibinfo {year} {1998})}\BibitemShut {NoStop}%
\bibitem [{\citenamefont {Davydov}\ \emph
  {et~al.}(2018{\natexlab{a}})\citenamefont {Davydov}, \citenamefont
  {Sveshnikov},\ and\ \citenamefont {Voronina}}]{Davydov2018a}%
  \BibitemOpen
  \bibfield  {author} {\bibinfo {author} {\bibfnamefont {A.}~\bibnamefont
  {Davydov}}, \bibinfo {author} {\bibfnamefont {K.}~\bibnamefont {Sveshnikov}},
  \ and\ \bibinfo {author} {\bibfnamefont {Y.}~\bibnamefont {Voronina}},\
  }\href {\doibase 10.1142/S0217751X18500045} {\bibfield  {journal} {\bibinfo
  {journal} {Int. J. Mod. Phys. A}\ }\textbf {\bibinfo {volume} {33}},\
  \bibinfo {pages} {1850004} (\bibinfo {year}
  {2018}{\natexlab{a}})}\BibitemShut {NoStop}%
\bibitem [{\citenamefont {Davydov}\ \emph
  {et~al.}(2018{\natexlab{b}})\citenamefont {Davydov}, \citenamefont
  {Sveshnikov},\ and\ \citenamefont {Voronina}}]{Davydov2018b}%
  \BibitemOpen
  \bibfield  {author} {\bibinfo {author} {\bibfnamefont {A.}~\bibnamefont
  {Davydov}}, \bibinfo {author} {\bibfnamefont {K.}~\bibnamefont {Sveshnikov}},
  \ and\ \bibinfo {author} {\bibfnamefont {Y.}~\bibnamefont {Voronina}},\
  }\href {\doibase 10.1142/S0217751X18500057} {\bibfield  {journal} {\bibinfo
  {journal} {Int. J. Mod. Phys. A}\ }\textbf {\bibinfo {volume} {33}},\
  \bibinfo {pages} {1850005} (\bibinfo {year}
  {2018}{\natexlab{b}})}\BibitemShut {NoStop}%
\bibitem [{\citenamefont {Sveshnikov}\ \emph
  {et~al.}(2019{\natexlab{a}})\citenamefont {Sveshnikov}, \citenamefont
  {Voronina}, \citenamefont {Davydov},\ and\ \citenamefont
  {Grashin}}]{Sveshnikov2019a}%
  \BibitemOpen
  \bibfield  {author} {\bibinfo {author} {\bibfnamefont {K.}~\bibnamefont
  {Sveshnikov}}, \bibinfo {author} {\bibfnamefont {Y.}~\bibnamefont
  {Voronina}}, \bibinfo {author} {\bibfnamefont {A.}~\bibnamefont {Davydov}}, \
  and\ \bibinfo {author} {\bibfnamefont {P.}~\bibnamefont {Grashin}},\ }\href
  {\doibase doi.org/10.1134/S0040577919030024} {\bibfield  {journal} {\bibinfo
  {journal} {Theor. Math. Phys.}\ }\textbf {\bibinfo {volume} {198}},\ \bibinfo
  {pages} {331} (\bibinfo {year} {2019}{\natexlab{a}})}\BibitemShut {NoStop}%
\bibitem [{\citenamefont {Sveshnikov}\ \emph
  {et~al.}(2019{\natexlab{b}})\citenamefont {Sveshnikov}, \citenamefont
  {Voronina}, \citenamefont {Davydov},\ and\ \citenamefont
  {Grashin}}]{Sveshnikov2019b}%
  \BibitemOpen
  \bibfield  {author} {\bibinfo {author} {\bibfnamefont {K.}~\bibnamefont
  {Sveshnikov}}, \bibinfo {author} {\bibfnamefont {Y.}~\bibnamefont
  {Voronina}}, \bibinfo {author} {\bibfnamefont {A.}~\bibnamefont {Davydov}}, \
  and\ \bibinfo {author} {\bibfnamefont {P.}~\bibnamefont {Grashin}},\ }\href
  {\doibase doi.org/10.1134/S0040577919040056} {\bibfield  {journal} {\bibinfo
  {journal} {Theor. Math. Phys.}\ }\textbf {\bibinfo {volume} {199}},\ \bibinfo
  {pages} {533} (\bibinfo {year} {2019}{\natexlab{b}})}\BibitemShut {NoStop}%
\bibitem [{\citenamefont {Sveshnikov}\ and\ \citenamefont
  {Voronina}(2022)}]{Sveshnikov2022}%
  \BibitemOpen
  \bibfield  {author} {\bibinfo {author} {\bibfnamefont {K.}~\bibnamefont
  {Sveshnikov}}\ and\ \bibinfo {author} {\bibfnamefont {Y.}~\bibnamefont
  {Voronina}},\ }\href@noop {} {\bibfield  {journal} {\bibinfo  {journal} {In
  preparation}\ } (\bibinfo {year} {2022})}\BibitemShut {NoStop}%
\bibitem [{\citenamefont {Itzykson}\ and\ \citenamefont
  {Zuber}(1980)}]{Itzykson1980}%
  \BibitemOpen
  \bibfield  {author} {\bibinfo {author} {\bibfnamefont {C.}~\bibnamefont
  {Itzykson}}\ and\ \bibinfo {author} {\bibfnamefont {J.-B.}\ \bibnamefont
  {Zuber}},\ }\href@noop {} {\emph {\bibinfo {title} {Quantum Field Theory}}}\
  (\bibinfo  {publisher} {McGraw-Hill},\ \bibinfo {year} {1980})\BibitemShut
  {NoStop}%
\bibitem [{\citenamefont {Fano}(1961)}]{Fano1961}%
  \BibitemOpen
  \bibfield  {author} {\bibinfo {author} {\bibfnamefont {U.}~\bibnamefont
  {Fano}},\ }\href {\doibase 10.1103/PhysRev.124.1866} {\bibfield  {journal}
  {\bibinfo  {journal} {Phys. Rev.}\ }\textbf {\bibinfo {volume} {124}},\
  \bibinfo {pages} {1866} (\bibinfo {year} {1961})}\BibitemShut {NoStop}%
\bibitem [{\citenamefont {Voronina}\ \emph {et~al.}(2019)\citenamefont
  {Voronina}, \citenamefont {Sveshnikov}, \citenamefont {Grashin},\ and\
  \citenamefont {Davydov}}]{Voronina2019a}%
  \BibitemOpen
  \bibfield  {author} {\bibinfo {author} {\bibfnamefont {Y.}~\bibnamefont
  {Voronina}}, \bibinfo {author} {\bibfnamefont {K.}~\bibnamefont
  {Sveshnikov}}, \bibinfo {author} {\bibfnamefont {P.}~\bibnamefont {Grashin}},
  \ and\ \bibinfo {author} {\bibfnamefont {A.}~\bibnamefont {Davydov}},\ }\href
  {\doibase https://doi.org/10.1016/j.physe.2018.08.013} {\bibfield  {journal}
  {\bibinfo  {journal} {Physica E}\ }\textbf {\bibinfo {volume} {106}},\
  \bibinfo {pages} {298 } (\bibinfo {year} {2019})}\BibitemShut {NoStop}%
\bibitem [{\citenamefont {{Rajaraman}}(1982)}]{Rajaraman1982}%
  \BibitemOpen
  \bibfield  {author} {\bibinfo {author} {\bibfnamefont {R.}~\bibnamefont
  {{Rajaraman}}},\ }\href
  {https://inis.iaea.org/search/search.aspx?orig_q=RN:15036991} {\emph
  {\bibinfo {title} {Solitons and Instantons}}},\ \bibinfo {edition} {1st}\
  ed.\ (\bibinfo  {publisher} {North-Holland Publishing Company},\ \bibinfo
  {year} {1982})\BibitemShut {NoStop}%
\bibitem [{\citenamefont {Sveshnikov}(1991)}]{Sveshnikov1991}%
  \BibitemOpen
  \bibfield  {author} {\bibinfo {author} {\bibfnamefont {K.}~\bibnamefont
  {Sveshnikov}},\ }\href {\doibase 10.1016/0370-2693(91)90244-K} {\bibfield
  {journal} {\bibinfo  {journal} {Phys. Lett. B}\ }\textbf {\bibinfo {volume}
  {255}},\ \bibinfo {pages} {255} (\bibinfo {year} {1991})}\BibitemShut
  {NoStop}%
\bibitem [{\citenamefont {Sundberg}\ and\ \citenamefont
  {Jaffe}(2004)}]{Jaffe2004}%
  \BibitemOpen
  \bibfield  {author} {\bibinfo {author} {\bibfnamefont {P.}~\bibnamefont
  {Sundberg}}\ and\ \bibinfo {author} {\bibfnamefont {R.~L.}\ \bibnamefont
  {Jaffe}},\ }\href {\doibase 10.1016/j.aop.2003.08.015} {\bibfield  {journal}
  {\bibinfo  {journal} {Ann. Phys.}\ }\textbf {\bibinfo {volume} {309}},\
  \bibinfo {pages} {442} (\bibinfo {year} {2004})}\BibitemShut {NoStop}%
\bibitem [{\citenamefont {Zeldovich}\ and\ \citenamefont
  {Popov}(1972)}]{Zeldovich1972}%
  \BibitemOpen
  \bibfield  {author} {\bibinfo {author} {\bibfnamefont {Y.~B.}\ \bibnamefont
  {Zeldovich}}\ and\ \bibinfo {author} {\bibfnamefont {V.~S.}\ \bibnamefont
  {Popov}},\ }\href {http://stacks.iop.org/0038-5670/14/i=6/a=R01} {\bibfield
  {journal} {\bibinfo  {journal} {Soviet Physics Uspekhi}\ }\textbf {\bibinfo
  {volume} {14}},\ \bibinfo {pages} {673} (\bibinfo {year} {1972})}\BibitemShut
  {NoStop}%
\bibitem [{\citenamefont {Marsman}\ and\ \citenamefont
  {Horbatsch}(2011)}]{Marsman2011}%
  \BibitemOpen
  \bibfield  {author} {\bibinfo {author} {\bibfnamefont {A.}~\bibnamefont
  {Marsman}}\ and\ \bibinfo {author} {\bibfnamefont {M.}~\bibnamefont
  {Horbatsch}},\ }\href {\doibase 10.1103/PhysRevA.84.032517} {\bibfield
  {journal} {\bibinfo  {journal} {Phys. Rev. A}\ }\textbf {\bibinfo {volume}
  {84}},\ \bibinfo {pages} {032517} (\bibinfo {year} {2011})}\BibitemShut
  {NoStop}%
\bibitem [{\citenamefont {Maltsev}\ \emph {et~al.}(2020)\citenamefont
  {Maltsev}, \citenamefont {Shabaev}, \citenamefont {Zaytsev},\ and\
  \citenamefont {et~al.}}]{Maltsev2020a}%
  \BibitemOpen
  \bibfield  {author} {\bibinfo {author} {\bibfnamefont {I.}~\bibnamefont
  {Maltsev}}, \bibinfo {author} {\bibfnamefont {V.}~\bibnamefont {Shabaev}},
  \bibinfo {author} {\bibfnamefont {V.}~\bibnamefont {Zaytsev}}, \ and\
  \bibinfo {author} {\bibnamefont {et~al.}},\ }\href {\doibase
  10.1134/S0030400X2008024X} {\bibfield  {journal} {\bibinfo  {journal} {Opt.
  Spectrosc.}\ }\textbf {\bibinfo {volume} {128}},\ \bibinfo {pages} {1100}
  (\bibinfo {year} {2020})}\BibitemShut {NoStop}%
\bibitem [{\citenamefont {Gumberidze}\ \emph {et~al.}(2009)\citenamefont
  {Gumberidze}, \citenamefont {St{\"o}hlker}, \citenamefont {Beyer},
  \citenamefont {Bosch}, \citenamefont {Bräuning-Demian}, \citenamefont
  {Hagmann}, \citenamefont {Kozhuharov}, \citenamefont {K\"uhl}, \citenamefont
  {Mann}, \citenamefont {Indelicato}, \citenamefont {Quint}, \citenamefont
  {Schuch},\ and\ \citenamefont {Warczak}}]{FAIR2009}%
  \BibitemOpen
  \bibfield  {author} {\bibinfo {author} {\bibfnamefont {A.}~\bibnamefont
  {Gumberidze}}, \bibinfo {author} {\bibfnamefont {T.}~\bibnamefont
  {St{\"o}hlker}}, \bibinfo {author} {\bibfnamefont {H.~F.}\ \bibnamefont
  {Beyer}}, \bibinfo {author} {\bibfnamefont {F.}~\bibnamefont {Bosch}},
  \bibinfo {author} {\bibfnamefont {A.}~\bibnamefont {Bräuning-Demian}},
  \bibinfo {author} {\bibfnamefont {S.}~\bibnamefont {Hagmann}}, \bibinfo
  {author} {\bibfnamefont {C.}~\bibnamefont {Kozhuharov}}, \bibinfo {author}
  {\bibfnamefont {T.}~\bibnamefont {K\"uhl}}, \bibinfo {author} {\bibfnamefont
  {R.}~\bibnamefont {Mann}}, \bibinfo {author} {\bibfnamefont {P.}~\bibnamefont
  {Indelicato}}, \bibinfo {author} {\bibfnamefont {W.}~\bibnamefont {Quint}},
  \bibinfo {author} {\bibfnamefont {R.}~\bibnamefont {Schuch}}, \ and\ \bibinfo
  {author} {\bibfnamefont {A.}~\bibnamefont {Warczak}},\ }\href {\doibase
  https://doi.org/10.1016/j.nimb.2008.10.079} {\bibfield  {journal} {\bibinfo
  {journal} {Nucl. Instr. {\&} Meth. in Phys. Research B}\ }\textbf {\bibinfo
  {volume} {267}},\ \bibinfo {pages} {248} (\bibinfo {year}
  {2009})}\BibitemShut {NoStop}%
\bibitem [{\citenamefont {Ter-Akopian}\ \emph {et~al.}(2015)\citenamefont
  {Ter-Akopian}, \citenamefont {Greiner}, \citenamefont {Meshkov},
  \citenamefont {Oganessian}, \citenamefont {Reinhardt},\ and\ \citenamefont
  {Trubnikov}}]{Ter2015}%
  \BibitemOpen
  \bibfield  {author} {\bibinfo {author} {\bibfnamefont {G.~M.}\ \bibnamefont
  {Ter-Akopian}}, \bibinfo {author} {\bibfnamefont {W.}~\bibnamefont
  {Greiner}}, \bibinfo {author} {\bibfnamefont {I.}~\bibnamefont {Meshkov}},
  \bibinfo {author} {\bibfnamefont {Y.}~\bibnamefont {Oganessian}}, \bibinfo
  {author} {\bibfnamefont {J.}~\bibnamefont {Reinhardt}}, \ and\ \bibinfo
  {author} {\bibfnamefont {G.}~\bibnamefont {Trubnikov}},\ }\href {\doibase
  10.1142/S0218301315500160} {\bibfield  {journal} {\bibinfo  {journal} {Int.
  J. Mod. Phys. E}\ }\textbf {\bibinfo {volume} {24}},\ \bibinfo {pages}
  {1550016} (\bibinfo {year} {2015})}\BibitemShut {NoStop}%
\bibitem [{\citenamefont {Ma}\ \emph {et~al.}(2017)\citenamefont {Ma},
  \citenamefont {Wen}, \citenamefont {Zhang}, \citenamefont {Yu}, \citenamefont
  {Cheng}, \citenamefont {Yang}, \citenamefont {Huang}, \citenamefont {Wang},
  \citenamefont {Zhu}, \citenamefont {Cai}, \citenamefont {Zhao}, \citenamefont
  {Mao}, \citenamefont {Yang}, \citenamefont {Zhou}, \citenamefont {Xu},
  \citenamefont {Yuan}, \citenamefont {Xia}, \citenamefont {Zhao},
  \citenamefont {Xiao},\ and\ \citenamefont {Zhan}}]{MA2017169}%
  \BibitemOpen
  \bibfield  {author} {\bibinfo {author} {\bibfnamefont {X.}~\bibnamefont
  {Ma}}, \bibinfo {author} {\bibfnamefont {W.}~\bibnamefont {Wen}}, \bibinfo
  {author} {\bibfnamefont {S.}~\bibnamefont {Zhang}}, \bibinfo {author}
  {\bibfnamefont {D.}~\bibnamefont {Yu}}, \bibinfo {author} {\bibfnamefont
  {R.}~\bibnamefont {Cheng}}, \bibinfo {author} {\bibfnamefont
  {J.}~\bibnamefont {Yang}}, \bibinfo {author} {\bibfnamefont {Z.}~\bibnamefont
  {Huang}}, \bibinfo {author} {\bibfnamefont {H.}~\bibnamefont {Wang}},
  \bibinfo {author} {\bibfnamefont {X.}~\bibnamefont {Zhu}}, \bibinfo {author}
  {\bibfnamefont {X.}~\bibnamefont {Cai}}, \bibinfo {author} {\bibfnamefont
  {Y.}~\bibnamefont {Zhao}}, \bibinfo {author} {\bibfnamefont {L.}~\bibnamefont
  {Mao}}, \bibinfo {author} {\bibfnamefont {J.}~\bibnamefont {Yang}}, \bibinfo
  {author} {\bibfnamefont {X.}~\bibnamefont {Zhou}}, \bibinfo {author}
  {\bibfnamefont {H.}~\bibnamefont {Xu}}, \bibinfo {author} {\bibfnamefont
  {Y.}~\bibnamefont {Yuan}}, \bibinfo {author} {\bibfnamefont {J.}~\bibnamefont
  {Xia}}, \bibinfo {author} {\bibfnamefont {H.}~\bibnamefont {Zhao}}, \bibinfo
  {author} {\bibfnamefont {G.}~\bibnamefont {Xiao}}, \ and\ \bibinfo {author}
  {\bibfnamefont {W.}~\bibnamefont {Zhan}},\ }\href {\doibase
  https://doi.org/10.1016/j.nimb.2017.03.129} {\bibfield  {journal} {\bibinfo
  {journal} {Nucl. Instr. {\&} Meth. in Phys. Research B}\ }\textbf {\bibinfo
  {volume} {408}},\ \bibinfo {pages} {169} (\bibinfo {year}
  {2017})}\BibitemShut {NoStop}%
\bibitem [{\citenamefont {M{\"u}ller-Nehler}\ and\ \citenamefont
  {Soff}(1994)}]{Mueller1994}%
  \BibitemOpen
  \bibfield  {author} {\bibinfo {author} {\bibfnamefont {U.}~\bibnamefont
  {M{\"u}ller-Nehler}}\ and\ \bibinfo {author} {\bibfnamefont {G.}~\bibnamefont
  {Soff}},\ }\href {\doibase 10.1016/0370-1573(94)90068-X} {\bibfield
  {journal} {\bibinfo  {journal} {Phys.Rep.}\ }\textbf {\bibinfo {volume}
  {246}},\ \bibinfo {pages} {101} (\bibinfo {year} {1994})}\BibitemShut
  {NoStop}%
\bibitem [{\citenamefont {Adkins}\ \emph {et~al.}(1983)\citenamefont {Adkins},
  \citenamefont {Nappi},\ and\ \citenamefont {Witten}}]{ADKINS1983552}%
  \BibitemOpen
  \bibfield  {author} {\bibinfo {author} {\bibfnamefont {G.~S.}\ \bibnamefont
  {Adkins}}, \bibinfo {author} {\bibfnamefont {C.~R.}\ \bibnamefont {Nappi}}, \
  and\ \bibinfo {author} {\bibfnamefont {E.}~\bibnamefont {Witten}},\ }\href
  {\doibase https://doi.org/10.1016/0550-3213(83)90559-X} {\bibfield  {journal}
  {\bibinfo  {journal} {Nucl. Phys. B}\ }\textbf {\bibinfo {volume} {228}},\
  \bibinfo {pages} {552} (\bibinfo {year} {1983})}\BibitemShut {NoStop}%
\bibitem [{\citenamefont {Holzwarth}\ and\ \citenamefont
  {Schwesinger}(1986)}]{Holzwarth_1986}%
  \BibitemOpen
  \bibfield  {author} {\bibinfo {author} {\bibfnamefont {G.}~\bibnamefont
  {Holzwarth}}\ and\ \bibinfo {author} {\bibfnamefont {B.}~\bibnamefont
  {Schwesinger}},\ }\href {\doibase 10.1088/0034-4885/49/8/001} {\bibfield
  {journal} {\bibinfo  {journal} {Reports on Progress in Physics}\ }\textbf
  {\bibinfo {volume} {49}},\ \bibinfo {pages} {825} (\bibinfo {year}
  {1986})}\BibitemShut {NoStop}%
\bibitem [{\citenamefont {Weigel}(2007)}]{Weigel2007}%
  \BibitemOpen
  \bibfield  {author} {\bibinfo {author} {\bibfnamefont {H.}~\bibnamefont
  {Weigel}},\ }\href@noop {} {\emph {\bibinfo {title} {Chiral Soliton Models
  for Baryons}}},\ Vol.\ \bibinfo {volume} {743 of Lecture Notes in Physics}\
  (\bibinfo  {publisher} {Springer},\ \bibinfo {year} {2007})\BibitemShut
  {NoStop}%
\bibitem [{\citenamefont {Hosaka}\ and\ \citenamefont
  {Toki}(2001)}]{Hosaka2001}%
  \BibitemOpen
  \bibfield  {author} {\bibinfo {author} {\bibfnamefont {A.}~\bibnamefont
  {Hosaka}}\ and\ \bibinfo {author} {\bibfnamefont {H.}~\bibnamefont {Toki}},\
  }\href@noop {} {\emph {\bibinfo {title} {Quarks, Baryons and Chiral
  Symmetry}}}\ (\bibinfo  {publisher} {World Scientific},\ \bibinfo {year}
  {2001})\BibitemShut {NoStop}%
\bibitem [{\citenamefont {Dubikovsky}\ and\ \citenamefont
  {Sveshnikov}(1994)}]{DUBIKOVSKY199480}%
  \BibitemOpen
  \bibfield  {author} {\bibinfo {author} {\bibfnamefont {A.}~\bibnamefont
  {Dubikovsky}}\ and\ \bibinfo {author} {\bibfnamefont {K.}~\bibnamefont
  {Sveshnikov}},\ }\href {\doibase
  https://doi.org/10.1016/0370-2693(94)90330-1} {\bibfield  {journal} {\bibinfo
   {journal} {Phys. Lett. B}\ }\textbf {\bibinfo {volume} {321}},\ \bibinfo
  {pages} {80} (\bibinfo {year} {1994})}\BibitemShut {NoStop}%
\bibitem [{\citenamefont {Weyl}(2015)}]{Weyl2015}%
  \BibitemOpen
  \bibfield  {author} {\bibinfo {author} {\bibfnamefont {H.}~\bibnamefont
  {Weyl}},\ }\href@noop {} {\emph {\bibinfo {title} {Symmetry}}},\ Vol.\
  \bibinfo {volume} {104 of Princeton Science Library}\ (\bibinfo  {publisher}
  {Princeton University Press},\ \bibinfo {year} {2015})\BibitemShut {NoStop}%
\end{thebibliography}%

\end{document}